\documentclass[a4paper,english]{aa}
\usepackage[T1]{fontenc}
\usepackage[latin1]{inputenc}
\usepackage{graphicx}
\usepackage{amssymb}
\IfFileExists{url.sty}{\usepackage{url}}
                      {\newcommand{\url}{\texttt}}
\usepackage[authoryear]{natbib}

\makeatletter

\newcommand{\noun}[1]{\textsc{#1}}

\providecommand{\tabularnewline}{\\}

\usepackage{txfonts}
\usepackage{aalongtable}
\bibpunct{(}{)}{;}{a}{}{,}

\usepackage{babel}
\makeatother
\begin{document}

\title{Spectrophotometric properties of galaxies at intermediate redshifts
(z\textasciitilde{}0.2--1.0)\thanks{Based on observations collected at the Very Large Telescope, European Southern Observatory, Paranal, Chile (ESO Programs 64.O-0439, 65.O-0367, 67.B-0255, 69.A-0358, and 72.A-0603)}}

\subtitle{I. Sample description, photometric properties and spectral measurements\thanks{Tables~\ref{photodata},\ref{classifdata},\ref{datalines1} and~\ref{datalines2} are only available in electronic form at the CDS via anonymous ftp to cdsarc.u-strasbg.fr (130.79.128.5) or via http://cdsweb.u-strasbg.fr/cgi-bin/qcat?J/A+A/}}

\author{F. Lamareille\inst{1} \and T. Contini\inst{1} \and J.-F. Le Borgne\inst{1} \and J.
Brinchmann\inst{2,3} \and S. Charlot\inst{2,4} \and J. Richard\inst{1} }

\institute{Laboratoire d'Astrophysique de Toulouse et Tarbes (LATT - UMR 5572),
Observatoire Midi-Pyrénées, 14 avenue E. Belin, F-31400 Toulouse,
France\and Max-Planck Institut für Astrophysik, Karl-Schwarzschild-Strasse
1 Postfach 1317, D-85741 Garching, Germany\and Centro de Astrof\'isica
da Universidade do Porto, Rua das Estrelas - 4150-762 Porto, Portugal\and Institut
d'Astrophysique de Paris, CNRS, 98 bis Boulevard Arago, F-75014 Paris,
France}

\offprints{F. Lamareille, e-mail: \texttt{flamare@ast.obs-mip.fr}}

\date{Received ; Accepted}

\abstract{We present the spectrophotometric properties of a sample of 141 emission-line
galaxies at redshifts in the range $0.2<z<1.0$ with a peak around
$z\in[0.2,0.4]$. The analysis is based on medium resolution ($R_{\mathrm{s}}=500-600$),
optical spectra obtained at VLT and Keck. The targets are mostly {}``Canada-France
Redshift Survey'' emission-line galaxies, with the addition of field
galaxies randomly selected behind lensing clusters. We complement
this sample with galaxy spectra from the {}``Gemini Deep Deep Survey''
public data release. We have computed absolute magnitudes of the galaxies
and measured the line fluxes and equivalent widths of the main emission/absorption
lines. The last two have been measured after careful subtraction of
the fitted stellar continuum using the \texttt{platefit} software
originally developed for the SDSS and adapted to our data. We present
a careful comparison of this software with the results of manual measurements.
The pipeline has also been tested on lower resolution spectra, typical
of the {}``VIMOS/VLT Deep Survey'' ($R_{\mathrm{s}}=250$), by resampling
our medium resolution spectra. We show that we can successfully deblend
the most important strong emission lines. These data are primarily
used to perform a spectral classification of the galaxies in order
to distinguish star-forming galaxies from AGNs. Among the initial
sample of 141 emission-line galaxies, we find 7 Seyfert 2 (narrow-line
AGN), 115 star-forming galaxies and 16 {}``candidate'' star-forming
galaxies. Scientific analysis of these data, in terms of chemical
abundances, stellar populations, etc, will be presented in subsequent
papers of this serie. }

\maketitle

\section{Introduction}

Understanding the major steps in the evolution of galaxies still remains
a great challenge to modern astrophysics. While the general theoretical
framework of the hierarchical growth of structures in the universe
including the build up of galaxies is well in place, this picture
remains largely unconstrained by observations, especially at high
redshifts. Statistically significant samples of galaxies, from the
local Universe to the highest redshifts, are crucial to constrain
the models of galaxy formation and evolution. Indeed, comparing the
physical properties (star formation rate, extinction, chemical abundances,
kinematics, mass, stellar populations, etc) of galaxies at different
epochs will allow us to study the evolution with redshift of fundamental
scaling relations such as the Luminosity-Metallicity or the Tully-Fisher
relations and hence put strong constraints on galaxy formation and
evolution models.

Thanks to recent massive surveys ({}``Sloan Digital Sky Survey''
SDSS, \citealt{Abazajian:2003AJ....126.2081A,Abazajian:2004AJ....128..502A};
{}``2degree Field Galaxy Redshift Survey'' 2dFGRS, \citealt{Colless:2001MNRAS.328.1039C}),
large spectroscopic samples of galaxies are now available in the local
universe, giving access to the detailed physical properties of galaxies
as a function of their environment for more than $\sim100\,000$ of
them. Similar massive spectroscopic surveys are being carried out
on the largest ground-based telescopes to explore the high-redshift
($0.2<z<4$) universe (e.g. {}``VIMOS/VLT Deep Survey'' VVDS, \citealt{LeFevre:2003SPIE.4834..173L};
{}``Deep Extragalactic Evolutionary Probe'' DEEP, \citealt{Koo:2002AAS...20113701K};
etc...). The main goal of these surveys is to study the evolution
of galaxies, large-scale structures, and Active Galactic Nuclei (AGN)
over more than 90\% of the current age of the universe (e.g. \citealp{LeFevre:2004A&A...417..839L}).
Most previous studies of intermediate-redshift ($z\sim0.2-1$) galaxies
have been driven by the {}``Canada-France Redshift Survey'' (CFRS,
\citealt{Lilly:1995ApJ...455...50L}) which produced a unique sample
of 591 field galaxies with $I_{\mathrm{AB}}<22.5$ in the range $0<z<1.4$
with a median redshift of $\sim0.56$ \citep{Lilly:1995ApJ...455...50L}.
Deep multicolor ($B,V,I$ and $K$) photometry is available for most
galaxies and several objects have been observed with the Hubble Space
Telescope (HST, \citealt{Brinchmann:1998ApJ...499..112B}) providing
useful complementary informations on the morphology \citep{Lilly:1998ApJ...500...75L,Schade:1999ApJ...525...31S}
and the level of interactions \citep{LeFevre:2000MNRAS.311..565L}
of galaxies up to $z\sim1$. This survey has been, for some years,
a unique tool for statistical studies of the evolution of field galaxies
as a function of redshift. However, because of the low spectral resolution
($\Delta\lambda\sim40$\AA) and limited signal-to-noise ratio (hereafter
SNR) of original CFRS spectra, no reliable estimate of crucial physical
properties, such as chemical abundances and reddening, have been determined
from these data.

Subsequent spectrophotometric studies of intermediate-redshift galaxies
have been performed by various authors on smaller samples (e.g. \citealp{Guzman:1997ApJ...489..559G,Kobulnicky:1999ApJ...511..118K,Hammer:2001ApJ...550..570H,Contini:2002MNRAS.330...75C,Lilly:2003ApJ...597..730L,Kobulnicky:2003ApJ...599.1006K,Liang:2004A&A...423..867L,Liang:2004A&A...417..905L,Maier:2004A&A...418..475M}).
Most of these studies were focused on galaxies with either a peculiar
morphology (e.g. compact and luminous galaxies; \citealt{Guzman:1997ApJ...489..559G,Hammer:2001ApJ...550..570H})
or selected in a special wavelength domain: UV-bright \citep{Contini:2002MNRAS.330...75C},
infrared-bright \citep{Liang:2004A&A...423..867L}, or narrow-band
selected galaxies \citep{Maier:2004A&A...418..475M}.

Thanks to the new class of multi-object spectrograph and to the associated
large and deep surveys (VVDS, DEEP, etc), a large amount of spectrophotometric
data will now become available. One of the goals of this paper is
to review all the technical issues that can be involved in the reduction
and analysis process of these large datasets, together with the scientific
results that can be drawn from these studies. This will allow us to
define a standard pipeline with particular care taken to optimise
it for the VVDS.

This paper builds up on previous work dedicated to the spectrophotometric
analysis of SDSS data (e.g. \citealp{Tremonti:2004astro.ph..5537T,Brinchmann:2004MNRAS.351.1151B}),
in which a large part of the pipeline has been described already.
In this paper, we describe how we adapt the existing pipeline to the
study of intermediate-redshift galaxies observed at a lower spectral
resolution and SNR than the SDSS galaxies. In order to do that, we
defined a sample of $\sim140$ galaxies at intermediate redshifts
($0.2<z<1.0$) showing a large range of physical properties. Using
medium-resolution optical spectra mainly acquired with the FORS (FOcal
Reducer Spectrograph) instruments on the VLT, we derived their spectrophotometric
properties, with a particular attention on defining an automatic process
which will be mandatory to analyze large surveys. We also investigate
the effect of spectral resolution on the derived quantities, as large
surveys like VVDS are based on low-resolution spectra.

This first paper focuses on the general reduction pipeline, photometric
properties and emission-line measurements. The scientific analysis
of this sample in terms of stellar populations, chemical abundances,
etc, will be presented in subsequent papers.

This paper is organized as follow: we first describe our sample in
Sect.~\ref{sec:Sample-description}, and then the observations and
associated data reduction in Sect.~\ref{sec:Data-reduction}. We
present the spectroscopic analysis in Sect.~\ref{sec:Spectroscopic-analysis}
and the photometric data in Sect.~\ref{sec:Photometric-analysis}.
Finally we perform a spectral classification of our sources in Sect.~\ref{sec:Spectral-classification}.

\section{Sample description\label{sec:Sample-description}}

\subsection{The parent samples}

The CFRS produced a large and homogeneous sample of field galaxies
with measured redshifts and morphological properties. This gives us
the opportunity to select interesting galaxies at intermediate redshifts
in order to acquire new spectra with a better spectral resolution
and SNR than the original ones. We thus decided to select and re-observe
a sub-sample of CFRS galaxies selected in three of the five CFRS fields
visible from Paranal (Chile), namely CFRS\,0000+00 (hereafter CFRS00),
CFRS\,0300+00 (hereafter CFRS03), and CFRS\,2215+00 (hereafter CFRS22).
In addition to this main sample, we acquired spectra for some new
and unidentified galaxies selected to fill the slits in the FORS masks.
This sample of 63 galaxies is called the {}``CFRS sub-sample'' (see
Table~\ref{cfsample}).

In addition, we decided to take advantage of some series of spectra
previously observed by the {}``Galaxies'' team in Toulouse and their
collaborators. They were essentially samples of galaxies inside massive
lensing clusters, but, in order to complete the masks, some foreground
or background field galaxies were observed. These 48 field galaxies
form the {}``CLUST sub-sample'' (see Table~\ref{clsample}).

Finally, we added a sample of public available spectra from the {}``Gemini
Deep Deep Survey'' (GDDS, \citealt{Abraham:2004AJ....127.2455A})
to cover the high redshift end (i.e. $0.4<z<1.0$). We selected 31
emission-line spectra which form the {}``GDDS sub-sample'' (see
Table~\ref{gdsample}).

\begin{table}

\caption{The CFRS sub-sample. \emph{(a)} Unique identification number (this
work, please use the acronym {}``LCL05'', for reference). \emph{(b)}
CFRS id if available. \emph{(c)} Redshift (see Sect.~\ref{sub:Redshifts}
for redshift determination). \emph{(d)} Signal-to-noise ratio of the
continuum at $\sim5500\AA$ or $\sim3500\AA$ (noted by the symbol
$^{\textrm{*}}$). \emph{(e)} Maximum signal-to-noise ratio of the
strongest emission line. \emph{(f)} 'a' flag means the emission lines
were manually measured on this spectrum, 'c' flag means this spectrum
was a combination of two observations.}

\label{cfsample}

{\scriptsize \begin{tabular}{lllrrrr}
\hline\hline
LCL05$^{\textrm{a}}$	&CFRS$^{\textrm{b}}$	&J2000 ($\alpha,\delta$)	&$z^{\textrm{c}}$	&S/N$_c^{\textrm{d}}$	&S/N$_m^{\textrm{e}}$	&$^{\textrm{f}}$	\\
\hline
\hline
\multicolumn{7}{l}{field: CFRS00}\\
\hline
\object{001}	&	&$00~02~46.93~-00~39~01$	&$0.3405$	&$7.2$	&$17$	&a	\\
\object{002}	&	&$00~02~43.10~-00~40~48$	&$0.6157$	&$7.3^{\textrm{*}}$	&$23$	&a	\\
\object{003}	&	&$00~02~41.53~-00~40~01$	&$0.3409$	&$11.1$	&$58$	&a	\\
\object{004}	&\object{00.0852}	&$00~02~39.83~-00~41~02$	&$0.2682$	&$13.1$	&$50$	&a	\\
\object{005}	&\object{00.0861}	&$00~02~39.41~-00~41~40$	&$0.2682$	&$19.7$	&$19$	&	\\
\object{006}	&\object{00.0900}	&$00~02~37.06~-00~40~36$	&$0.2470$	&$8.7$	&$33$	&a	\\
\object{007}	&\object{00.0940}	&$00~02~35.58~-00~41~06$	&$0.2694$	&$13.5$	&$26$	&	\\
\object{008}	&\object{00.1013}	&$00~02~32.97~-00~41~33$	&$0.2437$	&$7.5$	&$34$	&	\\
\object{009}	&\object{00.0124}	&$00~02~29.91~-00~41~42$	&$0.2880$	&$8.5$	&$16$	&	\\
\object{010}	&\object{00.0148}	&$00~02~28.19~-00~41~16$	&$0.2672$	&$14.0$	&$39$	&c	\\
\object{011}	&\object{00.1726}	&$00~02~48.51~-00~41~35$	&$0.2959$	&$16.1$	&$55$	&a	\\
\object{012}	&\object{00.0699}	&$00~02~44.91~-00~41~23$	&$0.0874$	&$11.6$	&$17$	&	\\
\object{013}	&	&$00~02~44.59~-00~39~52$	&$0.2489$	&$11.2$	&$21$	&a	\\
\object{014}	&	&$00~02~38.75~-00~40~21$	&$0.3902$	&$7.2$	&$7$	&a	\\
\object{015}	&\object{00.1057}	&$00~02~34.15~-00~41~31$	&$0.2432$	&$5.8$	&$30$	&a	\\
\object{016}	&\object{00.0121}	&$00~02~30.16~-00~41~35$	&$0.2975$	&$12.8$	&$228$	&a	\\
\object{017}	&\object{00.0229}	&$00~02~23.26~-00~41~23$	&$0.2453$	&$13.9$	&$18$	&	\\
\hline
\multicolumn{7}{l}{field: CFRS03}\\
\hline
\object{018}	&\object{03.1184}	&$03~02~49.28~+00~13~37$	&$0.2046$	&$7.3$	&$30$	&c	\\
\object{019}	&\object{03.1343}	&$03~02~49.56~+00~11~58$	&$0.1889$	&$5.0$	&$33$	&	\\
\object{020}	&\object{03.0442}	&$03~02~44.89~+00~13~45$	&$0.4781$	&$6.3^{\textrm{*}}$	&$27$	&c	\\
\object{021}	&\object{03.0476}	&$03~02~43.11~+00~14~13$	&$0.2601$	&$13.4$	&$73$	&ac	\\
\object{022}	&\object{03.0488}	&$03~02~42.16~+00~13~24$	&$0.6049$	&$5.6^{\textrm{*}}$	&$38$	&a	\\
\object{023}	&\object{03.0507}	&$03~02~40.44~+00~14~03$	&$0.4648$	&$6.9^{\textrm{*}}$	&$30$	&	\\
\object{024}	&\object{03.0523}	&$03~02~39.34~+00~13~27$	&$0.6532$	&$7.0^{\textrm{*}}$	&$47$	&ac	\\
\object{025}	&\object{03.0578}	&$03~02~35.19~+00~14~10$	&$0.2188$	&$7.4$	&$32$	&	\\
\object{026}	&\object{03.0605}	&$03~02~33.01~+00~14~07$	&$0.2189$	&$9.6$	&$54$	&ac	\\
\object{027}	&\object{03.0003}	&$03~02~31.85~+00~13~18$	&$0.2186$	&$4.3$	&$46$	&ac	\\
\object{028}	&\object{03.0037}	&$03~02~29.48~+00~14~13$	&$0.1744$	&$26.0$	&$86$	&ac	\\
\object{029}	&\object{03.0046}	&$03~02~28.67~+00~13~33$	&$0.5123$	&$5.5^{\textrm{*}}$	&$18$	&c	\\
\object{030}	&\object{03.0085}	&$03~02~25.24~+00~13~24$	&$0.6083$	&$4.1^{\textrm{*}}$	&$23$	&c	\\
\object{031}	&\object{03.0096}	&$03~02~24.29~+00~12~28$	&$0.2189$	&$7.8$	&$60$	&ac	\\
\hline
\multicolumn{7}{l}{field: CFRS22}\\
\hline
\object{032}	&\object{22.0502}	&$22~17~58.26~+00~14~29$	&$0.4682$	&$6.0$	&$15$	&	\\
\object{033}	&\object{22.0585}	&$22~17~55.60~+00~16~59$	&$0.2940$	&$11.9$	&$32$	&c	\\
\object{034}	&\object{22.0671}	&$22~17~53.03~+00~18~27$	&$0.3175$	&$15.5$	&$67$	&ac	\\
\object{035}	&\object{22.0819}	&$22~17~48.76~+00~17~18$	&$0.2910$	&$10.4$	&$36$	&ac	\\
\object{036}	&\object{22.0855}	&$22~17~47.88~+00~16~28$	&$0.2105$	&$13.6$	&$100$	&a	\\
\object{037}	&\object{22.0975}	&$22~17~45.12~+00~14~47$	&$0.4189$	&$6.8$	&$21$	&	\\
\object{038}	&\object{22.1013}	&$22~17~44.31~+00~15~05$	&$0.2307$	&$14.4$	&$76$	&a	\\
\object{039}	&\object{22.1084}	&$22~17~42.53~+00~14~21$	&$0.2928$	&$16.3$	&$52$	&ac	\\
\object{040}	&\object{22.1203}	&$22~17~39.54~+00~15~25$	&$0.5384$	&$9.5^{\textrm{*}}$	&$69$	&c	\\
\object{052}	&\object{22.0474}	&$22~17~58.70~+00~21~11$	&$0.2794$	&$8.0$	&$119$	&a	\\
\object{053}	&\object{22.0504}	&$22~17~58.07~+00~21~37$	&$0.5382$	&$10.9 ^{\textrm{*}}$	&$66$	&a	\\
\object{054}	&\object{22.0637}	&$22~17~54.01~+00~21~26$	&$0.5422$	&$16.4 ^{\textrm{*}}$	&$83$	&a	\\
\object{055}	&\object{22.0642}	&$22~17~53.77~+00~22~05$	&$0.4687$	&$3.7^{\textrm{*}}$	&$10$	&a	\\
\object{056}	&\object{22.0717}	&$22~17~51.63~+00~21~46$	&$0.2787$	&$19.6$	&$38$	&	\\
\object{057}	&\object{22.0823}	&$22~17~48.57~+00~21~27$	&$0.3333$	&$24.6$	&$28$	&	\\
\object{058}	&\object{22.1082}	&$22~17~42.49~+00~21~05$	&$0.2918$	&$4.7$	&$71$	&	\\
\object{059}	&	&$22~17~44.00~+00~23~21$	&$0.2765$	&$8.6$	&$28$	&a	\\
\object{060}	&\object{22.1144}	&$22~17~40.75~+00~21~46$	&$0.3586$	&$5.0$	&$21$	&a	\\
\object{061}	&\object{22.1220}	&$22~17~38.82~+00~21~19$	&$0.3583$	&$8.9$	&$29$	&a	\\
\object{062}	&\object{22.1231}	&$22~17~38.42~+00~22~13$	&$0.2846$	&$16.8$	&$82$	&a	\\
\object{063}	&\object{22.1309}	&$22~17~36.18~+00~21~24$	&$0.2847$	&$5.2$	&$23$	&	\\
\object{041}	&	&$22~17~53.01~+00~19~14$	&$0.2164$	&$2.7$	&$9$	&	\\
\object{042}	&	&$22~17~53.48~+00~19~25$	&$0.3524$	&$2.2$	&$22$	&a	\\
\object{043}	&\object{22.0622}	&$22~17~54.58~+00~16~58$	&$0.3237$	&$6.4$	&$20$	&	\\
\object{044}	&	&$22~17~46.54~+00~17~13$	&$0.2764$	&$19.9$	&$24$	&	\\
\object{045}	&\object{22.0919}	&$22~17~46.48~+00~16~53$	&$0.4712$	&$6.1^{\textrm{*}}$	&$158$	&ac	\\
\object{046}	&	&$22~17~46.99~+00~16~23$	&$0.6515$	&$7.3^{\textrm{*}}$	&$35$	&	\\
\object{047}	&	&$22~17~47.12~+00~16~26$	&$0.4716$	&$2.8^{\textrm{*}}$	&$24$	&	\\
\object{048}	&\object{22.0903}	&$22~17~46.76~+00~15~45$	&$0.2948$	&$4.6$	&$22$	&c	\\
\object{049}	&\object{22.0832}	&$22~17~48.44~+00~15~15$	&$0.2306$	&$25.1$	&$91$	&ac	\\
\object{050}	&\object{22.1064}	&$22~17~43.08~+00~15~08$	&$0.5369$	&$4.4^{\textrm{*}}$	&$42$	&ac	\\
\object{051}	&\object{22.1339}	&$22~17~35.39~+00~14~34$	&$0.3842$	&$6.1^{\textrm{*}}$	&$78$	&ac	\\
\hline
\end{tabular}
}
\end{table}
\begin{table}

\caption{The CLUST sub-sample. Same legend as Table~\ref{cfsample}. \emph{(g)}
alternative identification number if available (LBP2003: \citealt{LeBorgne:2003A&A...402..433L},
CBB2001: \citealt{Couch:2001ApJ...549..820C}, CPK2001: \citealt{Campusano:2001A&A...378..394C},
SKK2001: \citealt{Smail:2001MNRAS.323..839S}).}

\label{clsample} 

{\scriptsize \begin{tabular}{lllrrr}
\hline\hline
LCL05$^{\textrm{a}}$	&alt$^{\textrm{g}}$	&J2000 ($\alpha,\delta$)	&$z^{\textrm{c}}$	&S/N$_c^{\textrm{d}}$	&S/N$_m^{\textrm{e}}$	\\
\hline
\hline
\multicolumn{5}{l}{field: a2218}\\
\hline
\object{136}	&\object{SKK2001 368}	&$16~35~59.12~+66~12~01.3$	&$0.6926$	&$4.0^{\textrm{*}}$	&$17$	\\
\object{137}	&\object{SKK2001 159}	&$16~35~45.02~+66~12~44.7$	&$0.4730$	&$5.7^{\textrm{*}}$	&$24$	\\
\object{138}	&	&$16~35~40.48~+66~13~06.0$	&$0.4491$	&$1.4$	&$6$	\\
\hline
\multicolumn{5}{l}{field: a2390}\\
\hline
\object{064}	&	&$21~53~38.10~+17~43~48.0$	&$0.2412$	&$13.6$	&$70$	\\
\object{065}	&	&$21~53~40.01~+17~44~07.2$	&$0.0665$	&$6.1$	&$214$	\\
\object{066}	&	&$21~53~28.00~+17~39~01.1$	&$0.4261$	&$1.0$	&$8$	\\
\object{067}	&	&$21~53~25.34~+17~39~44.4$	&$0.4500$	&$3.3$	&$10$	\\
\object{068}	&	&$21~53~30.42~+17~39~16.4$	&$0.6291$	&$2.6^{\textrm{*}}$	&$11$	\\
\object{069}	&	&$21~53~26.84~+17~40~43.4$	&$0.2213$	&$6.8$	&$38$	\\
\object{070}	&	&$21~53~29.30~+17~40~26.8$	&$0.7392$	&$6.1^{\textrm{*}}$	&$27$	\\
\object{071}	&	&$21~53~33.45~+17~40~53.2$	&$0.5263$	&$3.9$	&$29$	\\
\object{072}	&	&$21~53~39.42~+17~43~50.6$	&$0.3425$	&$4.8$	&$15$	\\
\object{141}	&	&$21~53~33.02~+17~41~56.8$	&$0.3982$	&$5.7$	&$33$	\\
\hline
\multicolumn{5}{l}{field: a963}\\
\hline
\object{139}	&	&$10~17~04.82~+39~02~27.2$	&$0.7307$	&$1.2$	&$27$	\\
\object{140}	&	&$10~17~04.57~+39~02~25.3$	&$0.7307$	&$2.4$	&$25$	\\
\hline
\multicolumn{5}{l}{field: ac114}\\
\hline
\object{073}	&\object{LBP2003 b}	&$22~58~37.19~-34~49~27.8$	&$0.2605$	&$10.6$	&$42$	\\
\object{074}	&	&$22~58~43.42~-34~48~04.8$	&$0.0965$	&$13.7$	&$253$	\\
\object{075}	&\object{CBB2001 796}	&$22~58~54.75~-34~48~26.8$	&$0.0985$	&$10.2$	&$28$	\\
\object{076}	&\object{LBP2003 h}	&$22~58~43.35~-34~49~36.5$	&$0.3207$	&$21.5$	&$54$	\\
\object{077}	&\object{LBP2003 c}	&$22~58~43.07~-34~48~48.1$	&$0.2999$	&$19.8$	&$47$	\\
\object{078}	&\object{CPK2001 V7}	&$22~58~45.60~-34~49~03.9$	&$0.5669$	&$9.3^{\textrm{*}}$	&$27$	\\
\object{079}	&\object{CPK2001 V6}	&$22~58~50.94~-34~47~26.5$	&$0.4095$	&$5.0^{\textrm{*}}$	&$16$	\\
\object{080}	&\object{CBB2001 688}	&$22~58~41.83~-34~49~06.1$	&$0.3304$	&$8.5$	&$34$	\\
\object{081}	&\object{CPK2001 V11}	&$22~58~57.46~-34~47~06.8$	&$0.3805$	&$2.5$	&$13$	\\
\object{082}	&\object{CPK2001	V9}	&$22~58~56.56~-34~46~58.6$	&$0.4121$	&$7.2^{\textrm{*}}$	&$17$	\\
\object{083}	&	&$22~58~54.94~-34~46~32.6$	&$0.7262$	&$11.7^{\textrm{*}}$	&$111$	\\
\object{084}	&\object{CBB2001 453}	&$22~58~37.48~-34~50~16.2$	&$0.4100$	&$4.2^{\textrm{*}}$	&$24$	\\
\object{085}	&	&$22~58~35.92~-34~49~26.9$	&$0.7186$	&$3.0^{\textrm{*}}$	&$20$	\\
\object{086}	&	&$22~58~35.20~-34~48~59.0$	&$0.4125$	&$2.6$	&$11$	\\
\object{087}	&	&$22~58~42.52~-34~49~26.8$	&$0.4092$	&$1.3^{\textrm{*}}$	&$9$	\\
\object{088}	&	&$22~58~41.11~-34~48~48.2$	&$0.7571$	&$4.8^{\textrm{*}}$	&$70$	\\
\hline
\multicolumn{5}{l}{field: cl1358}\\
\hline
\object{142}	&	&$13~59~48.33~+62~31~18.4$	&$0.4069$	&$1.6$	&$72$	\\
\hline
\multicolumn{5}{l}{field: cl2244}\\
\hline
\object{089}	&	&$22~47~14.75~-02~03~25.1$	&$0.5628$	&$5.8^{\textrm{*}}$	&$23$	\\
\object{090}	&	&$22~47~14.63~-02~08~12.9$	&$0.5651$	&$5.4^{\textrm{*}}$	&$34$	\\
\object{091}	&	&$22~47~13.62~-02~07~36.5$	&$0.7865$	&$7.8^{\textrm{*}}$	&$27$	\\
\object{092}	&	&$22~47~08.58~-02~07~05.4$	&$0.3289$	&$4.6$	&$57$	\\
\object{093}	&	&$22~47~08.35~-02~06~38.8$	&$0.6402$	&$5.1^{\textrm{*}}$	&$36$	\\
\object{094}	&	&$22~47~07.04~-02~04~28.6$	&$0.5701$	&$2.4^{\textrm{*}}$	&$17$	\\
\object{095}	&	&$22~47~09.56~-02~07~15.7$	&$0.3416$	&$3.9$	&$13$	\\
\object{096}	&	&$22~47~14.16~-02~06~51.8$	&$0.4386$	&$3.3^{\textrm{*}}$	&$15$	\\
\object{097}	&	&$22~47~11.35~-02~06~29.6$	&$0.5724$	&$4.8^{\textrm{*}}$	&$17$	\\
\object{098}	&	&$22~47~11.08~-02~06~19.5$	&$0.5717$	&$2.9^{\textrm{*}}$	&$15$	\\
\hline
\multicolumn{5}{l}{field: j1206}\\
\hline
\object{099}	&	&$12~06~13.71~-08~51~01.3$	&$0.3555$	&$7.9$	&$21$	\\
\object{100}	&	&$12~06~10.29~-08~45~53.7$	&$0.3547$	&$5.6$	&$26$	\\
\object{101}	&	&$12~06~09.66~-08~50~44.4$	&$0.3547$	&$8.6$	&$18$	\\
\object{102}	&	&$12~06~10.97~-08~50~22.3$	&$0.4280$	&$10.0$	&$13$	\\
\object{103}	&	&$12~06~13.18~-08~48~26.8$	&$0.4759$	&$6.6$	&$21$	\\
\object{104}	&	&$12~06~07.74~-08~47~21.2$	&$0.4522$	&$5.2^{\textrm{*}}$	&$19$	\\
\hline
\end{tabular}
}
\end{table}
\begin{table}

\caption{The GDDS sub-sample. Same legend as Table~\ref{cfsample}.}

\label{gdsample} 

{\scriptsize \begin{tabular}{llrr}
\hline\hline
LCL05$^{\textrm{a}}$	&GDDS id	&J2000 ($\alpha,\delta$)	&$z$\\
\hline
\hline
\multicolumn{4}{l}{field: NOAO-Cetus}\\
\hline
\object{105}	&\object{02-0452}	&$02~09~49.51~-04~40~24.49$	&$0.828$	\\
\object{106}	&\object{02-0585}	&$02~09~50.13~-04~40~07.55$	&$0.825$	\\
\object{107}	&\object{02-0756}	&$02~09~43.49~-04~39~43.11$	&$0.864$	\\
\object{108}	&\object{02-0995}	&$02~09~48.09~-04~38~54.39$	&$0.786$	\\
\object{109}	&\object{02-1134}	&$02~09~44.46~-04~38~33.46$	&$0.913$	\\
\object{110}	&\object{02-1724}	&$02~09~37.13~-04~36~02.61$	&$0.996$	\\
\hline
\multicolumn{4}{l}{field: NTT Deep}\\
\hline
\object{111}	&\object{12-5337}	&$12~05~18.75~-07~24~57.19$	&$0.679$	\\
\object{112}	&\object{12-5513}	&$12~05~16.62~-07~24~43.70$	&$0.611$	\\
\object{113}	&\object{12-5685}	&$12~05~15.21~-07~24~28.16$	&$0.960$	\\
\object{114}	&\object{12-5722}	&$12~05~20.96~-07~24~22.27$	&$0.841$	\\
\object{115}	&\object{12-6456}	&$12~05~19.15~-07~23~45.64$	&$0.612$	\\
\object{116}	&\object{12-6800}	&$12~05~18.14~-07~23~21.97$	&$0.615$	\\
\object{117}	&\object{12-7099}	&$12~05~26.34~-07~22~53.02$	&$0.567$	\\
\object{118}	&\object{12-7205}	&$12~05~15.47~-07~22~58.00$	&$0.568$	\\
\object{119}	&\object{12-7660}	&$12~05~26.83~-07~22~07.83$	&$0.791$	\\
\object{120}	&\object{12-7939}	&$12~05~31.39~-07~20~37.77$	&$0.664$	\\
\object{121}	&\object{12-8250}	&$12~05~17.24~-07~20~02.97$	&$0.767$	\\
\hline
\multicolumn{4}{l}{field: SA22}\\
\hline
\object{122}	&\object{22-0040}	&$22~17~32.22~+00~12~45.91$	&$0.818$	\\
\object{123}	&\object{22-0145}	&$22~17~47.08~+00~13~17.40$	&$0.754$	\\
\object{124}	&\object{22-0563}	&$22~17~36.84~+00~15~27.22$	&$0.787$	\\
\object{125}	&\object{22-0619}	&$22~17~45.85~+00~16~42.48$	&$0.673$	\\
\object{126}	&\object{22-0630}	&$22~17~32.36~+00~16~16.28$	&$0.753$	\\
\object{127}	&\object{22-0643}	&$22~17~38.32~+00~16~59.41$	&$0.788$	\\
\object{128}	&\object{22-0751}	&$22~17~46.55~+00~16~26.68$	&$0.471$	\\
\object{129}	&\object{22-0926}	&$22~17~31.36~+00~17~48.10$	&$0.786$	\\
\object{130}	&\object{22-1534}	&$22~17~37.87~+00~17~45.88$	&$0.470$	\\
\object{131}	&\object{22-1674}	&$22~17~49.22~+00~17~14.32$	&$0.879$	\\
\object{132}	&\object{22-2196}	&$22~17~44.16~+00~15~21.56$	&$0.627$	\\
\object{133}	&\object{22-2491}	&$22~17~37.66~+00~14~12.38$	&$0.471$	\\
\object{134}	&\object{22-2541}	&$22~17~32.94~+00~13~58.92$	&$0.617$	\\
\object{135}	&\object{22-2639}	&$22~17~46.70~+00~13~31.93$	&$0.883$	\\
\hline
\end{tabular}
}
\end{table}

\subsection{Selection criteria}

The main goal of our program is to probe the physical properties of
star-forming galaxies at intermediate redshifts. We thus selected,
among the CFRS sub-sample, galaxies with narrow emission lines as
quoted in the literature, thus excluding galaxies with broad Balmer
emission lines typical of AGN. In order to obtain spectra with a sufficient
SNR in a reasonable exposure time, we limited ourselves to galaxies
brighter than an apparent $V$-band magnitude $V_{\mathrm{AB}}=23$
(on the CFRS sub-sample). In order to fill the MOS masks, some galaxies
without emission lines were also observed. Although we will not include
these objects in the present analysis, their spectra have been reduced
for possible future use. After the basic data reduction process (see
Sect.~\ref{sec:Data-reduction}), we selected only the spectra with
{}``visible'' (i.e. from visual examination, signal-to-noise ratio
of at least $~5$) emission lines and a good overall SNR of the continuum
(at least 10). We also want the spectrum to show at least {[}O\noun{ii}{]}$\lambda$3727,
H$\beta$ and {[}O\noun{iii}{]}$\lambda$5007 lines in order to
derive the metallicity of the galaxies.

We do not aim to construct any volume-limited, magnitude-limited,
or emission-line flux-limited sample. Our main concern is to build
a sample of star-forming galaxies selected by their bright emission
lines. However, we must point out that this selection criterion introduces
some biases. First the very high or very low metallicity objects will
not be selected (i.e. {[}O\noun{iii}{]} lines are too weak). Second,
very dusty and thus very strongly reddened galaxies are not selected
in our sample.

\section{Spectroscopic data\label{sec:Data-reduction}}

\subsection{Observations and data reduction}

Spectrophotometric observations of the {}``CFRS sub-sample'' were
performed during two observing runs (periods P65 and P67) with the
ESO/VLT at Paranal (Chile). Two nights (July 1st and August 28th,
2000) were devoted to the first run (ESO 65.O-0367) during which we
observed three masks: two in the CFRS22 field and one in the CFRS00
field. We used the FORS1 spectrograph mounted on the ANTU unit of
the VLT. The exposure time for each mask was divided into four exposures
of 40mn, leading to a total exposure time per mask of 2h40mn. Two
other nights (June 25th and September 13th, 2001) were allocated for
the second run (ESO 67.B-0255). For this run, we used both the FORS1
and FORS2 spectrograph mounted on the ANTU and KUEYEN units of the
VLT respectively. We observed three more masks: one in the CFRS22
field (total exposure time = 8 $\times$ 25mn = 3h20mn), one in the
CFRS00 field (total exposure time = 6 $\times$ 25mn = 2h30mn), and
one in the CFRS03 field (total exposure time = 8 $\times$ 25mn =
3h20mn).

The instrumental configuration was the same for all the observations.
MOS masks have been produced using the FIMS software. Pre-images (5mn
exposure time in $r_{\mathrm{Gunn}}$ band) for each field have been
acquired for an accurate positioning and orientation of the MOS masks.
The GRIS300V grism has been used to cover a total possible wavelength
range of $\sim$ 4500--8500 \AA\ with a resolution $R_{\mathrm{s}}=500$.
The effective wavelength range depends on the position of the slit/galaxy
in the MOS mask, being shorter at the edges of the mask. The slit
width was 1\arcsec\  yielding a nominal resolution of $\sim15$
\AA. The GG435+31 light blocking filter was used to avoid any second-order
contamination in the red part of the spectrum.

Most spectra of the {}``CLUST sub-sample'' have been obtained during
the run ESO 072.A-0603 with FORS2 on VLT/KUEYEN dedicated to the observation
of background galaxies magnified by massive clusters. As the main
targets do not fill the whole masks, slits have been designed on cluster
and foreground galaxies, as well as background unmagnified galaxies.
The clusters observed were Abell\,2390, AC\,114 and Clg\,2244-02
(hereafter Cl2244). FORS2 in MXU mode has been used with the GRIS300V
grism and an order sorting filter GG375, allowing a useful wavelength
range from 4000 \AA\ to 8600 \AA, and yielding a wavelength resolution
of $R_{\mathrm{s}}=500$. The observations were made in service mode
between August 29th and September 3rd, 2003. For each cluster mask,
a total exposure time of $\sim$ 4h was obtained. A 1\arcsec\ slit
width was used for each slit. Similar spectra were obtained on April
11th 2002 during a visitor mode run (ESO 69.A-0358) on cluster MACS
J1206.2-0847 (hereafter J1206) with the FORS1 spectrograph on VLT/MELIPAL
(see Ebeling et al., in preparation). The GRIS300V grism and a 1''-width
slit were used, yielding a wavelength coverage between $\sim$ 4000
\AA\  and 8600 \AA, and a wavelength resolution of $R_{\mathrm{s}}=500$.
An order sorting filter GG375 was used. The total exposure time was
38mn. The additional AC\,114 data were obtained on October 5, 1999
during the run ESO 64.O-0439 with FORS1 on VLT/ANTU (UT1) telescope.
The same G300V and 1''-width slit were used. These observations were
also part of a program to study magnified background galaxies. The
wavelength coverage is \textasciitilde{}4000--8000 \AA\ and the
resulting resolution 500. Depending on the mask used, the exposure
times were 2h15mn, 1h30mn or 1h17mn (see \citealp{Campusano:2001A&A...378..394C}).

The remaining spectra in the {}``CLUST sub-sample'' (with LCL05\#
$\geq$ 136) are more magnified objects serendipitously found during
a long-slit search for Lyman-$\alpha$ emitters at high redshift along
the critical lines of the clusters Abell\,963, Abell\,2218, Abell\,2390
and Clg\,1358+62 \citep{Santos:2004ApJ...606..683S,Ellis:2001ApJ...560L.119E}.
The double-beam Low Resolution Imaging Spectrograph (LRIS, \citealt{Oke:1995PASP..107..375O})
was used on the Keck telescope with a 1''-width long and 175''-length
long slit, a 600-line grating blazed at $\lambda$ 7500 \AA\ (resolution
$\sim$ 3.0 \AA{}) for the red channel and a 300-line grism blazed
at 5000 \AA\ with a dichroic at 6800 \AA\ (resolution $\sim$
3.5--4.0 \AA{}) for the blue channel of the instrument. More details
on these observations are given in \citet{Santos:2004ApJ...606..683S}.

\begin{figure}
\begin{center}\includegraphics[%
  width=1.0\columnwidth,
  keepaspectratio]{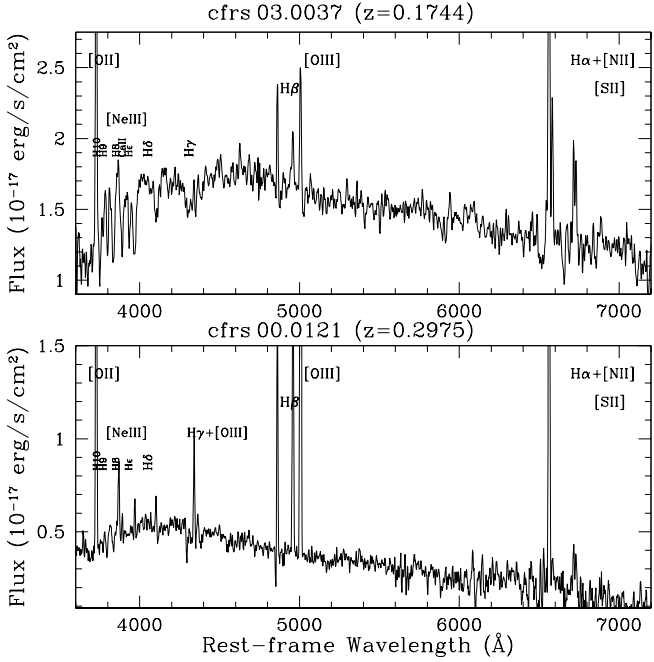}\end{center}

\caption{Examples of VLT/FORS spectra of intermediate-redshift CFRS galaxies.
Bottom panel: a low-metallicity galaxy (\object{CFRS\,00.0121})
with a high collisional excitation degree. Top panel: a high-metallicity
galaxy (\object{CFRS\,03.0037}). The position of the brightest
emission lines is indicated.}

\label{specs}
\end{figure}

Data reduction was performed in a standard way with IRAF packages.
In particular, the extraction of the 1D spectra and the computation
of SNR for each spectrum have been performed with the IRAF package
\texttt{apall}. The wavelength calibration used He-Ar arc lamps and
flux calibration have been done using spectrophotometric standard
stars observed each night. Two examples of FORS spectra of CFRS galaxies
are shown in Fig.~\ref{specs}.

Spectroscopic observations of the GDDS sub-sample have been done with
GMOS spectrograph on the Gemini North telescope between August 2002
and August 2003. The spectra cover a typical wavelength range of 5500
\AA{} to 9200 \AA{} with a wavelength resolution of approximately
$R_{\mathrm{s}}\approx630$ (see \citealt{Abraham:2004AJ....127.2455A}
for full details).

The spectroscopic observation details are summarized in Table~\ref{tabinstru}.

\begin{table*}

\caption{Summary of spectroscopic observations.}

\label{tabinstru}

{\scriptsize\begin{tabular}{lllllll}
\hline 
\hline
LCL05 ids&
instrument/telescope&
run&
$\lambda$ range (\AA)&
resolution&
slit width/lentgh&
exposure time\tabularnewline
\hline
001-010&
FORS1 / VLT&
65.O-0367&
4500-8500&
$R_{\mathrm{s}}=500$&
1'' / 22''&
4x40mn\tabularnewline
010-017&
FORS2 / VLT&
67.B-0255&
4500-8500&
$R_{\mathrm{s}}=500$&
1'' / 22''&
6x25mn\tabularnewline
018-031&
FORS1 / VLT&
67.B-0255&
4500-8500&
$R_{\mathrm{s}}=500$&
1'' / 22''&
2x25mn\tabularnewline
018-031(c)&
FORS2 / VLT&
67.B-0255&
4500-8500&
$R_{\mathrm{s}}=500$&
1'' / 22''&
6x25mn\tabularnewline
032-040&
FORS1 / VLT&
65.O-0367&
4500-8500&
$R_{\mathrm{s}}=500$&
1'' / 22''&
4x40mn\tabularnewline
033-040(c) / 042-051&
FORS2 / VLT&
67.B-0255&
4500-8500&
$R_{\mathrm{s}}=500$&
1'' / 22''&
6x25mn\tabularnewline
034-040(c) / 041 / 045-051(c)&
FORS1 / VLT&
67.B-0255&
4500-8500&
$R_{\mathrm{s}}=500$&
1'' / 22''&
2x25mn\tabularnewline
052-063&
FORS1 / VLT&
65.O-0367&
4500-8500&
$R_{\mathrm{s}}=500$&
1'' / 22''&
4x40mn\tabularnewline
064-072 / 083-098&
FORS2 / VLT&
72.A-0603&
4000-8600&
$R_{\mathrm{s}}=500$&
1'' / 22''&
$\sim$4h\tabularnewline
073-082&
FORS1 / VLT&
64.O-0439&
4000-8000&
$R_{\mathrm{s}}=500$&
1'' / 22''&
2h15mn, 1h30mn or 1h17mn\tabularnewline
099-104&
FORS1 / VLT&
69.A-0358&
4000-8600&
$R_{\mathrm{s}}=500$&
1'' / 22''&
38mn\tabularnewline
136 / 142&
LRIS / Keck&
2001A&
3800-10\,000&
$R_{\mathrm{s}}\approx2000$&
1'' / 175''&
33mn\tabularnewline
137-140&
LRIS / Keck&
2002A&
3800-10\,000&
$R_{\mathrm{s}}\approx2000$&
1'' / 175''&
33mn\tabularnewline
141&
LRIS / Keck&
2002B&
3800-10\,000&
$R_{\mathrm{s}}\approx2000$&
1'' / 175''&
33mn\tabularnewline
105-135&
GMOS / Gemini&
GDDS&
5500-9200&
$R_{\mathrm{s}}\approx630$&
&
\tabularnewline
\hline
\end{tabular}}
\end{table*}

\subsection{Redshift distribution\label{sub:Redshifts}}

The redshift of galaxies were derived using the centroid of the brightest
emission lines: {[}O\noun{ii}{]}$\lambda$3727, {[}O\noun{iii}{]}$\lambda$5007,
H$\beta$ and H$\alpha$ when available. In case of doubt, we tried
to adjust a stellar template to the continuum. Our redshifts agree
with the published ones to within 1\% for the re-observed CFRS galaxies.

\begin{figure}
\begin{center}\includegraphics[%
  width=1.0\columnwidth]{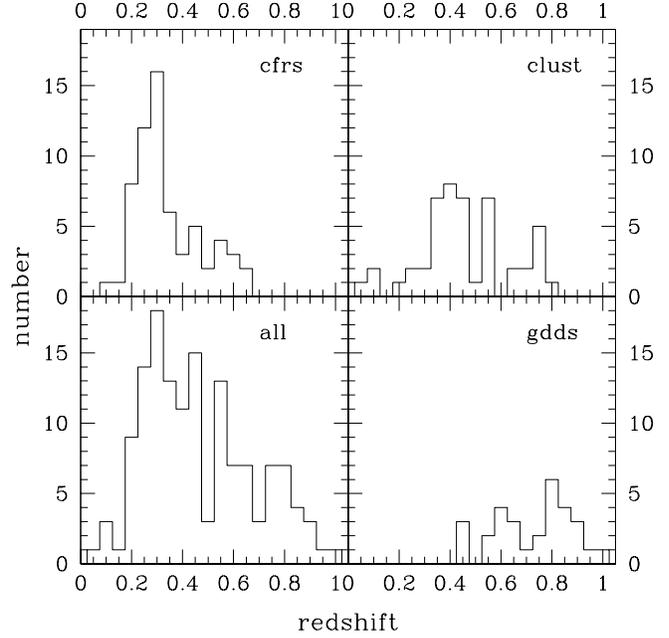}\end{center}

\caption{\label{histoz}Redshift histogram of our sample of intermediate-redshift
galaxies (bottom-left: all sample, top-left: {}``CFRS sub-samble'',
top-right: {}``CLUST sub-sample'', bottom-right: {}``GDDS sub-sample'').
The number of galaxies is calculated per 0.05 redshift bin.}
\end{figure}

Fig.~\ref{histoz} shows the histogram of the measured redshifts.
The redshift distribution is dominated by galaxies in the range $z\in[0.2,0.4]$.
This is a result of our selection criteria which favor galaxies showing
both {[}O\noun{ii}{]}$\lambda$3727 and H$\alpha$ emission lines.
This population is complemented by a number of galaxies with $z\in[0.4,1.0]$
leaving us with a statistically significant, although not complete,
sample of 141 galaxies spanning the redshift range 0.2 to 1.0.

\section{Spectroscopic analysis\label{sec:Spectroscopic-analysis}}

\subsection{Continuum fitting and subtraction}

\subsubsection{The software}

For the spectral fitting we have adapted the \texttt{platefit} IDL
code developed primarily by C.\ Tremonti. The code is discussed in
detail in \citet{Tremonti:2004astro.ph..5537T}, but for the benefit
of the reader we outline the key features here.

The continuum fitting is done by fitting a combination of model template
spectra (discussed below) to the observed spectrum with a non-negative
linear least squares fitting routine. The strong emission lines are
all masked out when carrying out this fit. The fitted continuum is
then subtracted from the object spectrum together with smoothed continuum
correction to take out minor spectrophotometric uncertainties. The
residual spectrum contains the emission lines.

The fit to the emission lines is carried out by fitting Gaussians
in velocity space to an adjustable list of lines. All forbidden lines
are tied to have the same velocity dispersion and all Balmer lines
are also tied together to have the same velocity dispersion. This
improves the fit for low SNR lines, but for the present sample this
is not of major importance. The weak {[}N\noun{ii}{]}$\lambda$6548
and {[}N\noun{ii}{]}$\lambda$6584 emission lines, which are closed
to the H$\alpha$ emission line at our working resolution, are tied
together so that the line ratio {[}N\noun{ii}{]}$\lambda$6584/{[}N\noun{ii}{]}$\lambda$6548
is equal to the theoretical value $3$. The {[}O\noun{ii}{]}$\lambda\lambda$3726,3729
line doublet is measured as one {[}O\noun{ii}{]}$\lambda$3727 emission
line, with a velocity dispersion freely fitted between $1.0$ and
$2.0$ times the velocity dispersion of the other forbidden lines,
which reproduces the broadening effect of two narrow lines blended
together. \texttt{platefit} returns the equivalent widths, fluxes
and associated errors for all fitted lines as well as other information.

\begin{figure}
\begin{center}\includegraphics[%
  width=1.0\columnwidth,
  keepaspectratio]{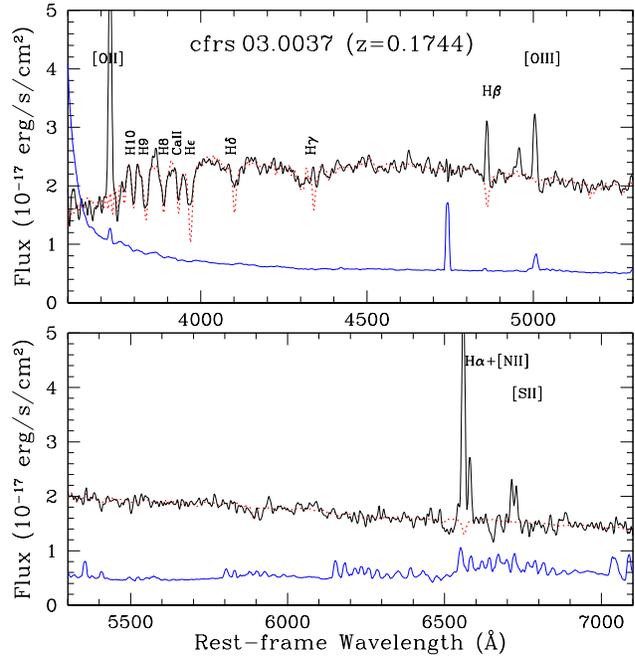}\end{center}

\caption{Example of input spectrum for the \object{CFRS\,03.0037} galaxy
at $z=0.1744$. The solid line shows the observed spectrum, the blue
line the error spectrum (magnified 5 times) and the red dotted line
shows the continuum fitting.}

\label{exspec1}
\end{figure}

The pipeline was optimised for SDSS spectra so some precautions must
be taken when using it on other data sets. In particular it is important
to have a reliable error estimate for each pixel (i.e. the error spectrum,
see Fig.~\ref{exspec1}) and to mask out regions of the spectra which
are unreliable. Failure to do so will severely affect the continuum
fitting.

\begin{figure}
\begin{center}\includegraphics[%
  width=1.0\columnwidth,
  keepaspectratio]{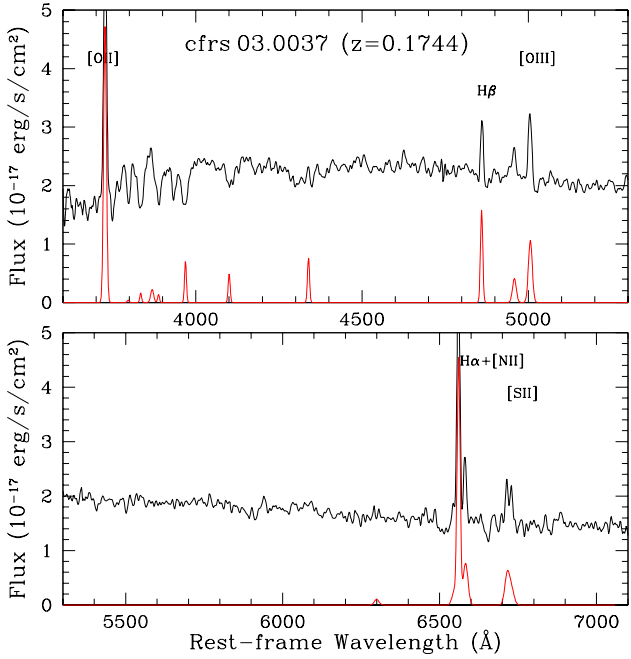}\end{center}

\caption{Example of emission-line fits, done after the continuum subtraction
for the \object{CFRS\,03.0037} galaxy at $z=0.1744$.}

\label{exspec2}
\end{figure}

The software returns a set of new spectra (sampled in velocity space):
the continuum spectrum which is the fitted linear combination of the
model templates added to the smoothed continuum (see Fig.~\ref{exspec1}),
the flux-continuum spectrum which is the raw spectrum with the stellar
continuum subtracted, the nebular spectrum which is built by adding
all the emission-line fits together (see Fig.~\ref{exspec2}), and
finally the stellar spectrum which is the raw spectrum with the nebular
one subtracted (note that this only take into account the lines which
are included in the fitting).


\subsubsection{Model templates}

The template spectra used to fit the continuum emission of the galaxy
in \texttt{platefit} were produced using the \citet{Bruzual:2003MNRAS.344.1000B}
population synthesis model%
\footnote{These template spectra are included in the original model release
package.%
}. At wavelengths between 3200 and 9500\,{}\AA{}, the template spectra
rely on the STELIB stellar spectral library \citep{LeBorgne:2003A&A...402..433L},
for which the resolution is about 3~\AA{} FWHM.

The template spectra were chosen in order to represent, through non-negative
linear combinations, the properties of galaxies with any star formation
history and metallicity. Specifically, the spectra were selected to
provide good coverage of SDSS-DR1 galaxies in the plane defined by
the 4000\,{}\AA{} break and the H$\delta$ stellar absorption line
strength, which are good indicators of the star formation history
of a galaxy (e.g., \citealt{Kauffmann:2003MNRAS.341...33K}). The
library includes 10 template spectra for each of the three metallicities
$Z=0.2\, Z_{\odot}$, $Z_{\odot}$ and $2.5\, Z_{\odot}$. The spectra
correspond to 10 instantaneous-burst models with ages of 0.005, 0.025,
0.10, 0.29, 0.64, 0.90, 1.4, 2.5, 5, and 11~Gyr.

\subsection{Adaptation to non-SDSS spectra and measurement of emission lines}

We have created an interface procedure which facilitates the analysis
of our non-SDSS spectra with the \texttt{platefit} routines. The input
spectra are provided as two FITS files each: one for the spectrum
itself and another one for the error spectrum, the output spectra
are written into ASCII files and the measurements are provided in
a FITS table. The behaviour of the interface procedure is controlled
by a parameter file which is an extension of that used in the \texttt{platefit}
code and which controls the operation of the code.

The result of flux and equivalent-width measurements of the main emission
lines is shown in Table~\ref{datalines1} and in Table~\ref{datalines2}.

\subsubsection{Comparison with manual determination}

It is instructive to compare the performance of the automatic fitting
code with manual measurements of line fluxes using standard methods.
To this end we measured emission lines from a subsample of the spectra
using the task \texttt{splot} in IRAF. This subsample is made of the
31 first reduced spectra, that do not show any specific properties,
among the CFRS sub-sample (see Table~\ref{cfsample}). In this section
we will compare these manual results to the automatically computed
ones. We expect to see significant differences for the Balmer lines
where it is difficult to adjust for the contribution of the underlying
stellar absorption when doing manual fitting. In contrast the measurements
for the forbidden lines should be consistent within the errors as
the effects of absorption lines for these is much less.

\begin{figure}
\begin{center}\includegraphics[%
  width=1.0\columnwidth]{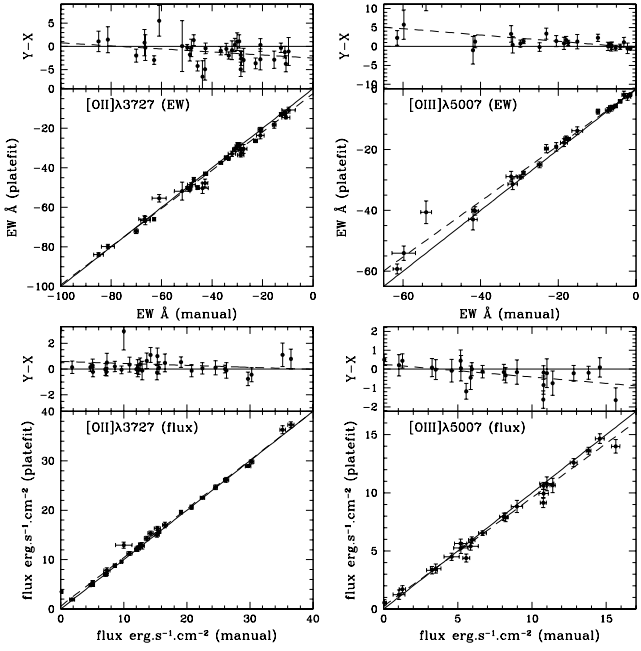} \end{center}

\caption{Comparison between oxygen emission lines measured automatically (using
\texttt{platefit}) and manually (using IRAF task \texttt{splot}).
Top panels: {[}O\noun{ii}{]}$\lambda$3727 (left) and {[}O\noun{iii}{]}$\lambda$5007
(right) equivalent widths (in \AA) given by \texttt{platefit} as
a function of the manual measurement. Bottom panels: same for the
measurements of line fluxes (in $10^{-17}$ erg s$^{-1}$ cm$^{-2}$).
The solid line is the $x=y$ line and the dashed line is the linear
regression.}

\label{figurecompare}
\end{figure}

In Figure~\ref{figurecompare}, we compare the automatic (using \texttt{platefit})
and manual measurements (using IRAF task \texttt{splot}) of oxygen
emission-line equivalent widths (top panels) and fluxes (bottom panels).
Figure~\ref{figurecompare} shows that there is a very good agreement
between manual and automatic measurements for two of the strongest
emission lines: {[}O\noun{ii}{]}$\lambda$3727 and {[}O\noun{iii}{]}$\lambda$5007.
Almost every point fall on the $y=x$ line and we also remark that
the error estimates are consistent between the two methods. By comparing
the bottom to the top panels, we see that the agreement is good both
for equivalent-width or for line-flux measurements.

\begin{figure}
\begin{center}\includegraphics[%
  width=1.0\columnwidth]{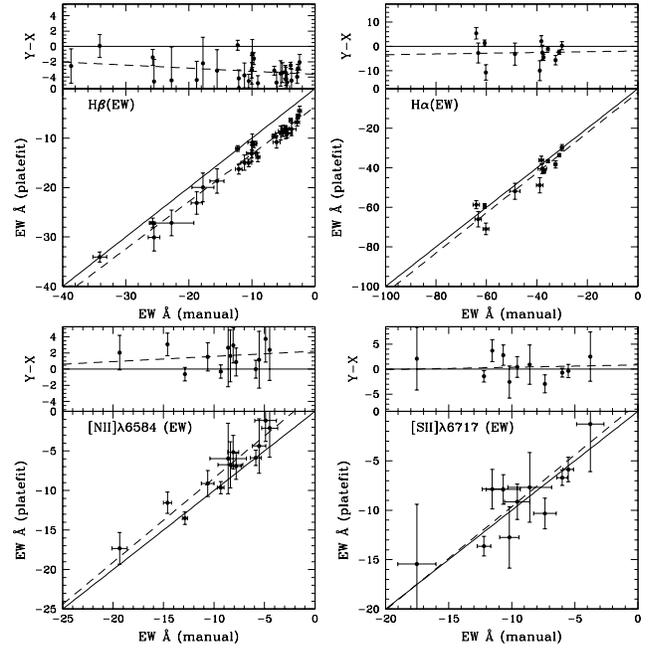}\end{center}

\caption{Comparison between equivalent widths (in \AA) for the Balmer (top
panels) and {[}N\noun{ii}{]}$\lambda$6584, {[}S\noun{ii}{]}$\lambda$6717
(bottom panels) emission lines measured automatically (using \texttt{platefit})
and manually (using IRAF task \texttt{splot}). The solid line is the
$x=y$ line and the dashed line is the linear regression.}

\label{figurecompare2}
\end{figure}

In Figure~\ref{figurecompare2}, we now compare the automatic and
manual measurements of equivalent widths for Balmer emission lines
(top panels) and low-intensity forbidden emission lines ({[}N\noun{ii}{]}$\lambda$6584
and {[}S\noun{ii}{]}$\lambda$6717, bottom panels). Figure~\ref{figurecompare2}
shows clearly the need to use \texttt{platefit} in order to have a
good estimate of the Balmer emission lines. As we see in the top-left
panel, manual measurements significantly underestimate the flux in
the H$\beta$ emission line where underlying stellar absorption is
normally not negligible in our galaxies. We note however that the
difference between manual and automatic measurements for the H$\alpha$
emission line is smaller, which is to be expected since the underlying
absorption is similar to that at H$\beta$, but the emission flux
is considerably higher. The bottom panels show the same comparison
for the fainter {[}N\noun{ii}{]}$\lambda$6584 and {[}S\noun{ii}{]}$\lambda$6717
emission lines. The dispersion here is larger but the measurements
are consistent within the errors.

\begin{figure}
\begin{center}\includegraphics[%
  width=1.0\columnwidth]{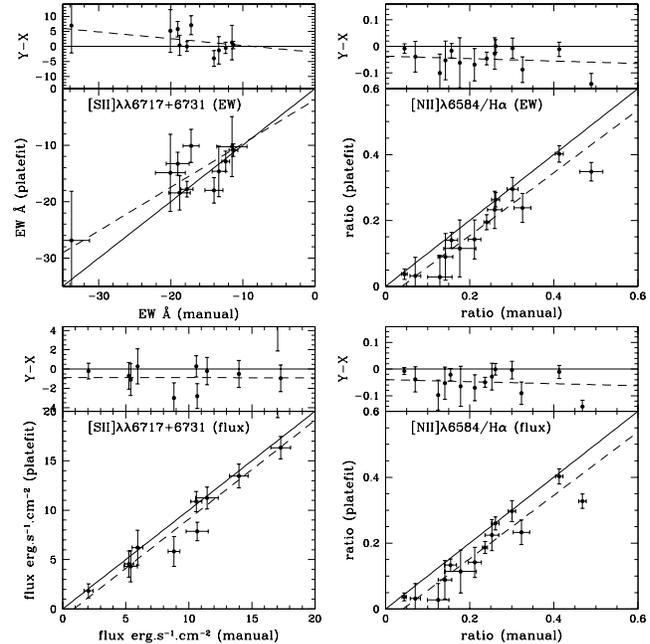}\end{center}

\caption{Comparison between blended features computed from equivalent width
(in \AA, top panels) and from fluxes (in $10^{-17}$ erg s$^{-1}$
cm$^{-2}$, bottom panels) automatically (using \texttt{platefit})
and manually (using IRAF task \texttt{splot}): the line sum {[}S\noun{ii}{]}$\lambda\lambda$6717+6731
(left) and the line ratio {[}N\noun{ii}{]}$\lambda$6584/H$\alpha$
(right). The solid line is the $x=y$ line and the dashed line is
the linear regression.}

\label{figurecompare3}
\end{figure}

Two features are based on two blended emission lines each at our working
resolution: the doublet {[}S\noun{ii}{]}$\lambda\lambda$6717+6731
and the line ratio {[}N\noun{ii}{]}$\lambda$6584/H$\alpha$. In
Figure~\ref{figurecompare3}, we compare the automatic and manual
measurements of these blended features (EWs: top panels, fluxes: bottom
panels). Figure~\ref{figurecompare3} illustrates the performance
of \texttt{platefit} in deblending these lines. We see that \texttt{platefit}
is able to give good results for the measurement of these low-intensity
blended emission lines. To reach this level of accuracy we had to
modify the way the equivalent width was estimated by \texttt{platefit},
which was optimised for higher resolution spectra. We tested various
methods, and found that the best results were obtained when we calculated
the equivalent width taking the continuum from the smoothed continuum
spectra and combined this with the emission line flux. This allows
us to make use of the line information in other parts of the spectrum
to overcome the blending problems and we get a very good agreement
between the measurements at different spectral resolutions as we will
see below.

\subsubsection{Resolution accuracy}

To prepare for the spectral analysis of upcoming deep surveys, such
as VVDS, we have used our medium resolution data to test the behaviour
of \texttt{platefit} when used on spectra with a lower resolution.
This point will be critical in particular for the {[}N\noun{ii}{]}$\lambda$6584/H$\alpha$
ratio, as these two lines are blended in low-resolution spectra ($R_{\mathrm{s}}\lesssim313$).
The main issue is to determine if we can use this ratio to perform
any spectral classification (see Sect.~\ref{sec:Spectral-classification})
and metallicity estimate \citep{VanZee:1998AJ....116.2805V,Pettini:2004MNRAS.348L..59P}.
Our sample is approximatively at the spectral resolution $R_{\mathrm{s}}=500$,
while the resolution of the VVDS is $R_{\mathrm{s}}=250$. Thus, we
have downgraded the resolution of our spectra by a factor of two with
a gaussian convolution, and we have rerun \texttt{platefit} on the
new spectra.

\begin{figure}
\begin{center}\includegraphics[%
  width=1.0\columnwidth]{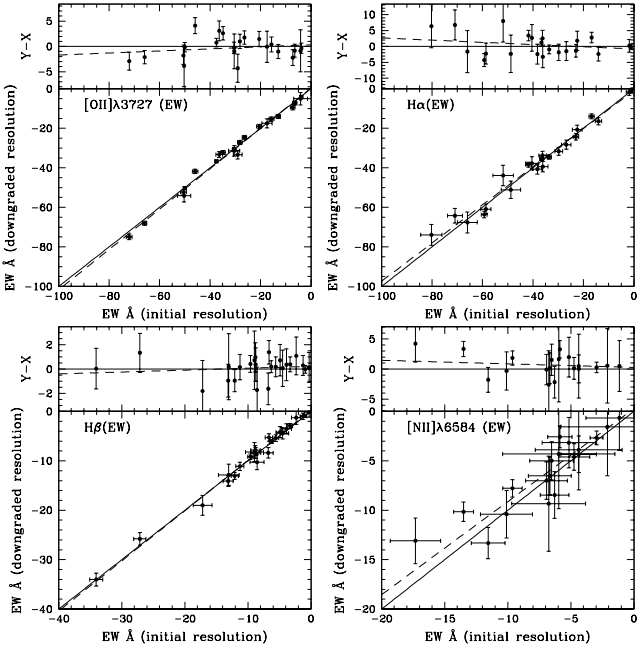}\end{center}

\caption{Comparison between the equivalent width (in \AA) of {[}O\noun{ii}{]}$\lambda$3727
(top-left), H$\alpha$ (top-right), H$\beta$ (bottom-left) and {[}N\noun{ii}{]}$\lambda$6584
(bottom-right) measured at downgraded ($R_{\mathrm{s}}=250$) and
original ($R_{\mathrm{s}}=500$) resolutions. The solid line is the
$x=y$ line and the dashed line is the linear regression.}

\label{figurecompvvds1}
\end{figure}

Table~\ref{tabcomparevvds} shows the difference between downgraded
resolution and original resolution measurements for some characteristic
lines. We see that the rms of the relative difference is low and strictly
less than the error associated on each line. We also remark that there
are some systematic shifts (i.e. the mean value of the difference
is not null) but they are still lower than the error. Figure~\ref{figurecompvvds1}
shows that there is no dependence with the line intensity. For low
resolution spectra, we reach a higher level of accuracy by tiding
up the velocity dispersion of all the emission lines together, whatever
they are forbidden or Balmer lines. This implies the assumption that
all broad-line AGNs have been taken out of the sample before running
the \texttt{platefit} software (see Sect.~\ref{sec:Spectral-classification}
below for a detailed discussion about the various spectral types of
emission-line galaxies).

\begin{figure}
\begin{center}\includegraphics[%
  width=1.0\columnwidth]{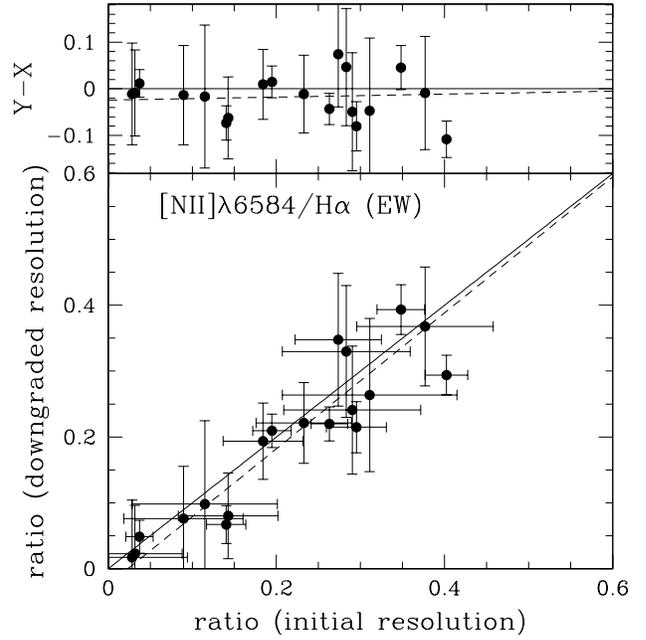}\end{center}

\caption{Comparison between the {[}N\noun{ii}{]}$\lambda$6584/H$\alpha$
line ratio calculated with the equivalent widths measured at downgraded
($R_{\mathrm{s}}=250$) and original ($R_{\mathrm{s}}=500$) resolutions.
The solid line is the $x=y$ line and the dashed line is the linear
regression.}

\label{figurecompvvds2}
\end{figure}

The {[}N\noun{ii}{]}$\lambda$6584/H$\alpha$ line ratio as measured
on the downgraded spectra is compared to the original measurements
in Fig.~\ref{figurecompvvds2} (see also Table~\ref{tabcomparevvds}).
It is clear that the difference is small and consistent with zero
within the errors. The logarithm of this line ratio, which is used
for metallicity estimates, also has a weak dependence on the spectral
resolution, and the scatter is lower than the standard error on metallicity
calibrations ($\sim0.2$ dex).

\begin{table}

\caption{Calculations of the mean and the rms of the difference between downgraded
resolution and original resolution measurements. We give first the
absolute values, and then the relative values which are more physically
significant in percent (\emph{rel.} columns). We compare these results
to the mean of the error on the original data (two last columns).}

\label{tabcomparevvds}

{\scriptsize\begin{tabular}{lllllll}
\hline 
\hline
parameter&
mean&
\emph{rel.}&
rms&
\emph{rel.}&
err&
\emph{rel.}\tabularnewline
\hline
EW({[}O\noun{ii}{]}$\lambda$3727)&
$-0.30$\AA&
$3.3$\%&
$2.09$\AA&
$9.8$\%&
$1.21$\AA&
$8.2$\%\tabularnewline
EW(H$\alpha$)&
$+0.51$\AA&
$0.5$\%&
$3.32$\AA&
$9.0$\%&
$1.95$\AA&
$9.2$\%\tabularnewline
EW({[}N\noun{ii}{]}$\lambda$6584)&
$+0.69$\AA&
$12$\%&
$1.92$\AA&
$24$\%&
$1.93$\AA&
$47$\%\tabularnewline
{[}N\noun{ii}{]}$\lambda$6584/H$\alpha$&
$-0.02$&
$11$\%&
$0.05$&
$23$\%&
$0.05$&
$47$\%\tabularnewline
log({[}N\noun{ii}{]}$\lambda$6584/H$\alpha$)&
$-0.06$ dex&
$8.2$\%&
$0.12$ dex&
$16$\%&
$0.20$ dex&
$21$\%\tabularnewline
\hline
\end{tabular}}
\end{table}


\section{Photometric analysis\label{sec:Photometric-analysis}}

\subsection{Photometric data}

We measured the photometric magnitudes with SExtractor \citep{Bertin:1996A&AS..117..393B}
in the $R$ band using the pre-imaging data. We used an input file
with all the image coordinates of the galaxies and we computed the
photometric magnitudes using the best radial adjustment (\texttt{MAG\_BEST}
parameter). We adopt these measurements in place of those from the
literature for the CFRS sub-sample to ensure consistency with the
CLUST sub-sample.

For the CFRS sub-sample, the pre-imaging was performed in the Gunn
$R$ band with the VLT/FORS1 camera. For the CLUST sub-sample, pre-images
have been acquired in the Bessel $R$ band with the FORS1 camera,
except for the J1206 field, for which the pre-imaging has been done
with the TEK2048 camera on the UH88in telescope in the $R$ band.
For the LRIS data, we used observations of Abell 963, Abell 2218,
Abell 2390 with the CFH12k camera at CFHT \citep{Czoske:2002astroph/0211517}
in the $I$ band. For the cluster Clg 1358+62, we measured photometry
on an HST-WFPC image in the $F606W$ band.

The photometric calibration was performed in different ways depending
on the field. The J1206 field was already calibrated. The other fields
from the CLUST sub-sample were calibrated using a photometric standard
star. For the CFRS sub-sample, the standard star was observed in a
different filter than the galaxies (Bessel $R$ rather than Gunn $R$),
preventing us from using it to do the calibration. Fortunately we
were able to take advantage of the previously measured magnitudes
of the CFRS galaxies in the $I$ band (CFHT FOCAM camera): the Bessel
$R$ magnitudes of these objects were computed using spectroscopic
colors (see Sect.~\ref{sub:Synthetic-magnitudes}), and we calculated
the zero-point of each image by doing a linear regression.

The photometric magnitudes of the GDDS sub-sample were directly taken
from the literature \citep{Chen:2002ApJ...570...54C}: $R$ band photometry
of the NTT Deep field has been taken with the BTC camera on the Cerro
Tololo Inter-American Observatory (CTIO) 4m telescope, $I$ band photometry
of the NOAO-Cetus and SA22 fields have been taken with the CFH12k
camera on the Canada France Hawaii Telescope (CFHT).

\subsection{Spectroscopic magnitudes\label{sub:Synthetic-magnitudes}}

We want to compute spectroscopic magnitudes, by integrating the flux
through a set of filter response curves, in order to have information
on the color (we only have photometric magnitudes in the $R$ band),
on the $k$-correction and on the aperture differences (i.e. the amount
of flux lost because of the limited size of the slits during spectroscopic
observations) of our galaxies.

We used an adaptation of the \texttt{filter\_thru} routine from the
SDSS IDL library%
\footnote{\url{http://spectro.princeton.edu}%
}. For a flux-calibrated spectrum this routine returns a spectroscopic
magnitude in the $AB$ system \citep{Oke:1974ApJS...27...21O}. If
the spectrum does not cover the full bandwidth of the filter (borders
at 5\%), it returns nothing. We computed spectroscopic magnitudes
directly from the observed spectrum if it covers the full bandwidth
of the filter, otherwise from the model spectrum given by the continuum
fitting.

We used the filter response curve of the FORS1 camera (CCD + atmosphere)
for the following bands: Bessel $B$, $V$, $R$ and $I$. We also
have the filter response of the FORS1 camera in the Gunn $R$ band
(used for the pre-imaging of the CFRS fields), of the TEK2048 camera
in the $R$ band (used for the pre-imaging of the J1206 field), of
the CFHT FOCAM camera in the $I$ band (used for the original CFRS
data), and of the BTC camera in the $R$ band and the CFH12k camera
in the $I$ band (used for the photometry of the GDDS sub-sample).
Finally we also calculated the photometry in the $u$, $g$, $r$,
and $i$ color system of the SDSS for possible comparison \citep{Fukugita:1996AJ....111.1748F}.
We have checked that the spectroscopic colors are in good agreement
(i.e. within the error bars) with published photometric colors.

We can use the information of the continuum SNR from Tables~\ref{cfsample}
and~\ref{clsample} to have an estimate of the uncertainties of the
spectroscopic magnitudes. We use the following formula:\[
\Delta m\approx\frac{2.5}{\ln10}\mathrm{SNR}^{-1}\]
We find a mean uncertainty of $\sim0.1$ magnitudes.

\subsection{Absolute magnitudes}

The absolute magnitudes were computed using photometric magnitudes
and the $k$-correction given from spectroscopic magnitudes (see Sect.~\ref{sub:Synthetic-magnitudes}).
If we want, for example, the absolute magnitude in the $I$ band ($M_{\mathrm{AB}}(I)$)
given a photometric magnitude in the $R$ band ($R_{\mathrm{AB}}$),
we use the following formula:\[
M_{\mathrm{AB}}(I)=d+R_{\mathrm{AB}}+\left(I_{\mathrm{spec}}^{\mathrm{rest}}-R_{\mathrm{spec}}^{\mathrm{obs}}\right)\]
 where $d$ is the distance modulus, $I_{\mathrm{spec}}^{\mathrm{rest}}$
and $R_{\mathrm{spec}}^{\mathrm{obs}}$ are the spectroscopic magnitudes
computed respectively in rest-frame and in observed-frame so that
$I_{\mathrm{spec}}^{\mathrm{rest}}-R_{\mathrm{spec}}^{\mathrm{obs}}$
is the $k$-correction. Note that we can alternatively write:\[
M_{\mathrm{AB}}(I)=d+I_{\mathrm{spec}}^{\mathrm{rest}}+\left(R_{\mathrm{AB}}-R_{\mathrm{spec}}^{\mathrm{obs}}\right)\]
 where $R_{\mathrm{AB}}-R_{\mathrm{spec}}^{\mathrm{obs}}$ is the
aperture difference. We remark that the aperture difference (\emph{ape}
in Table~\ref{photodata}) is less than $-1.3$ mag for a large majority
of our sample, which means that the aperture coverage is at least
$30$\% of the galaxy total luminosity. This avoids important disk/bulge
effects (see \citealp{Kewley:2005PASP..117..227K} for details). The
distance modulus is calculated using the last cosmology given by WMAP
\citep{Spergel:2003ApJS..148..175S}: $H_{\mathrm{0}}=71$ km s$^{-1}$
Mpc$^{-1}$, $\Omega_{\Lambda}=0.73$ and $\Omega_{\mathrm{m}}=0.27$.
The following formula gives the distance modulus as a function of
the redshift $z$:\[
d=-5\log\left(\frac{c}{H_{\mathrm{0}}\cdot10\textrm{pc}}\cdot(1+z)\cdot\int_{0}^{z}f(z')^{-1/2}dz'\right)\]

where $f(z')=(1+z')^{2}(1+\Omega_{\mathrm{m}}z')-\Omega_{\Lambda}z'(2+z')$.

The photometric magnitudes ($R_{AB}$ and $I_{AB}^{0}$ in Table~\ref{photodata})
are not corrected for foreground extinction whereas this is necessary
for future scientific analysis. We thus take into account the foreground
dust extinction from the Milky Way by using the \citet{Schlegel:1998ApJ...500..525S}
dust maps ($A_{I}$ in Table~\ref{photodata}) for computing the
rest-frame colors and the absolute magnitude.

\subsection{Lensing corrections}

For the galaxies in the CLUST sub-sample, we need to correct for the
magnification effect caused by the gravitational lensing of the cluster.
We do this using the most recent mass models for the galaxy clusters
in this sample (for AC\,114: \citealt{Natarajan:1998ApJ...499..600N,Campusano:2001A&A...378..394C};
for Abell\,2390: \citealt{Pello:1999A&A...346..359P}; for Cl\,2244:
Kneib et al., unpublished; for Abell\,2218: \citealt{Kneib:1996ApJ...471..643K,Ellis:2001ApJ...560L.119E};
for Abell\,963: \citealt{Smith:2003ApJ...590L..79S}; for Clg\,1358:\citealt{Franx:1997ApJ...486L..75F}).
We derived the magnification at the redshift of our background sources
with the LENSTOOL software developed by \citet{Kneib:1993PhDT.......189K}.
The corrections due to the lensing are usually small ($<0.3$ mags)
compared with the photometric errors. For sources very close to the
mean redshift of the cluster (for example in J\,1206) no correction
was applied.

The results are provided in Table~\ref{photodata}. The $I$-band
absolute magnitude is given after correction for the foreground dust
extinction, the $k$-correction and the lensing effect. We calculate
the absolute magnitude in any others bands using the given spectroscopic
colors.

\subsection{Color-color diagrams}

\begin{figure}
\begin{center}\includegraphics[%
  width=1.0\columnwidth]{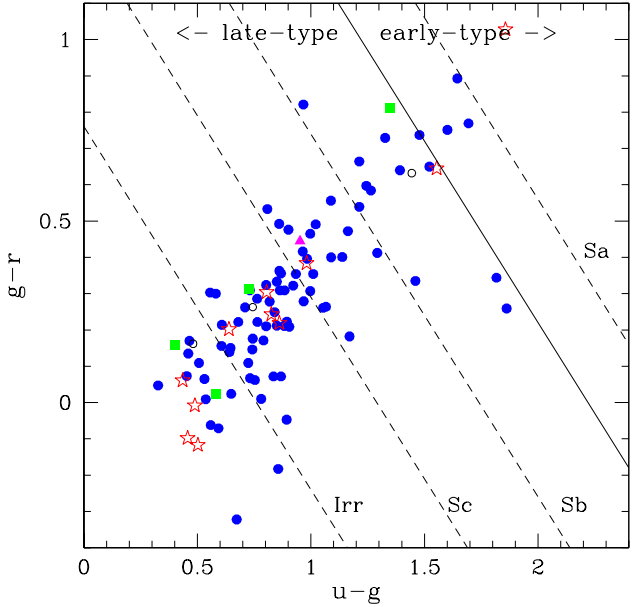}\end{center}

\caption{Rest-frame color-color diagram using SDSS passbands. The solid line
shows the empirical separation ($u-r=2.22$) between late-type (below
the line) and early-type (above the line) galaxies from \citet{SDSS:2001AJ....122.1861S}.
The dotted lines shows the mean value of the $u-r$ parameter for
standard spectra of the Sa, Sb, Sc and Irr spectral types: respectively
2.56, 1.74, 1.29 and 0.76 \citep{SDSS:2001AJ....122.1861S}. The different
symbols indicate star-forming galaxies and AGNs as determined in Sect.~\ref{sec:Spectral-classification}
(see the description of the symbols in Fig.~\ref{diagv32}).}

\label{color1}
\end{figure}

To gain some insight into the nature of the galaxy population in our
sample, we start by constructing the $g-r$ versus $u-g$ color-color
diagram (see Fig.~\ref{color1}). \citet{SDSS:2001AJ....122.1861S}
has shown that this diagram separates galaxies into early and late
types. We expect to see irregulars with a blue continuum at low $u-g$
and $g-r$ colors, whereas ellipticals, which have a substantial Balmer
break, should have {}``red'' $u-g$ and $g-r$ colours. To ease
interpretation we use rest-frame colours throughout.

\begin{figure}
\begin{center}\includegraphics[%
  width=1.0\columnwidth]{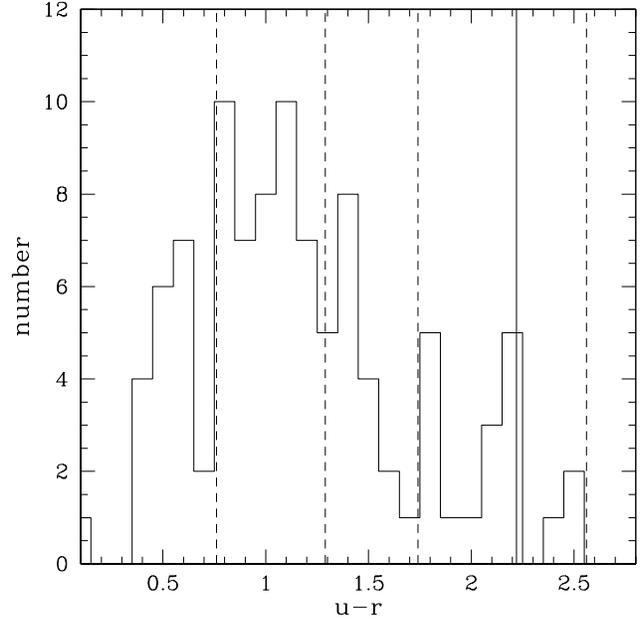}\end{center}

\caption{Histogram of the rest-frame $u-r$ color. We plot the number of galaxies
per 0.1 bin of the $u-r$ as an indicator of the spectral type. The
solid and dotted lines are the same as in Fig.~\ref{color1}.}

\label{color2}
\end{figure}

Figure~\ref{color1} shows that our galaxy sub-samples are well distributed
within the late-type region showing mainly irregular or Sc {}``color''
types. The proportion of early-type spirals is much less and we have
only a few ellipticals. This figure thus indicates that the latest
spectral types are more likely to be observed, as confirmed by the
histogram shown in Fig.~\ref{color2}. This result is primarily due
to our selection criterion as we biased our sample in favor of emission-line
galaxies. Indeed irregular galaxies usually have brighter emission
lines than Sb galaxies, so they will be in our sample down to very
low SNR. In contrast galaxies of (spectral) type Sb will only be included
in our sample when their spectrum is of good SNR. Any possible effect
introduced by this bias will have to be taken into account in subsequent
analysis.

We must however remark that neither Fig.~\ref{color1} nor Fig.~\ref{color2}
are accurate enough to determine which galaxy is of a given spectral
type because of the high dispersion of the $u-r$ values for each
spectral type (e.g. the effect of internal dust on the colors).

\section{Spectral classification\label{sec:Spectral-classification}}

\subsection{Nature of the main ionizing source}

As we want to focus the scientific analysis on star-forming galaxies,
we have to make the difference between starbursts and AGNs which both
show emission lines in their spectrum. The AGN population can be divided
into three main types: Seyfert 1, Seyfert 2 and LINERs. The Seyfert
1, also called broad-line AGNs, can be excluded from our sample by
comparing the FWHM of the Balmer emission lines to the FWHM of the
forbidden lines: those galaxies with a significantly higher FWHM for
the Balmer lines are expected to be Seyfert 1. The ratio of the FWHM
of the Balmer lines to that of the forbidden lines is consistent with
unity for most of our sample galaxies (mean of $0.98$ with a rms
scatter of $0.18$). We found 6 peculiar objects showing Balmer lines
significantly broader than forbidden ones ($FWHM_{\mathrm{Balmer}}/FWHM_{\mathrm{forbidden}}\sim2-3$).
These objects could be classified as Seyfert 1 galaxies. However,
after a careful visual inspection of individual spectra, we found
that the measurement of the FWHM of the Balmer lines in these galaxies
is disrupted by either weak Balmer emission lines or noise features.
These objects are thus classified as narrow emission-line galaxies.

\subsection{Diagnostic diagrams}

\subsubsection{{}``Red'' diagnostics diagrams}

We still need to separate star-forming galaxies from narrow-line AGNs,
namely Seyfert 2 and LINERs. Seyfert 2 have a high excitation degree
compared to LINERs. The standard prescription \citep{Veilleux:1987ApJS...63..295V}
makes use of the {[}N\noun{ii}{]}$\lambda$6584/H$\alpha$, {[}S\noun{ii}{]}$\lambda\lambda$6717+6731/H$\alpha$
and {[}O\noun{iii}{]}$\lambda$5007/H$\beta$ line ratios to separate
the star-forming galaxies from AGNs; and the {[}O\noun{iii}{]}$\lambda$5007/H$\beta$
as an indicator of the ionization level to distinguish Seyfert 2 from
LINERs.

\begin{figure}
\begin{center}\includegraphics[%
  width=1.0\columnwidth,
  keepaspectratio]{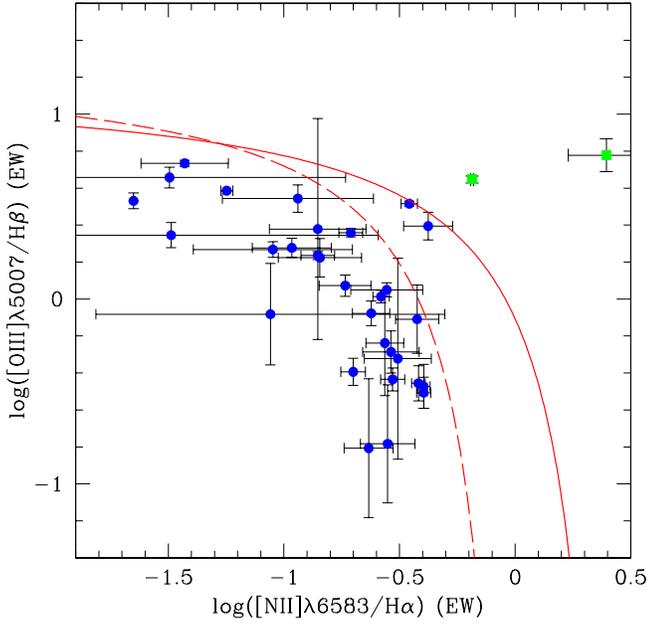}\end{center}

\caption{Standard diagnostic diagram using the {[}N\noun{ii}{]}$\lambda$6584/H$\alpha$
emission-line ratio. Blue circles identify star-forming galaxies,
green squares indicate AGNs (Seyfert 2) and magenta triangles show
contradictory cases (see text for details). The solid line shows the
theoretical limit from \citet{Kewley:2001ApJS..132...37K}. The dashed
line is the empirical limit from \citet{Kauffmann:2003MNRAS.346.1055K}
for SDSS galaxies.}

\label{diagn}
\end{figure}

\begin{figure}
\begin{center}\includegraphics[%
  width=1.0\columnwidth,
  keepaspectratio]{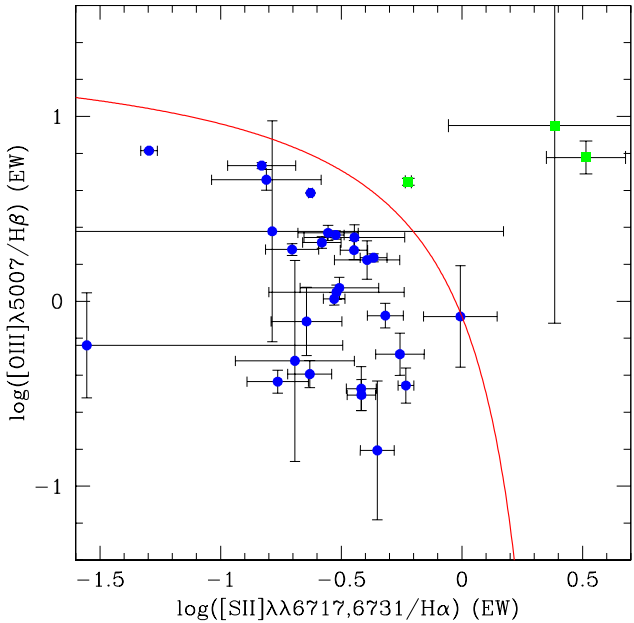}\end{center}

\caption{Standard diagnostic diagram using the {[}S\noun{ii}{]}$\lambda\lambda$6717+6731/H$\alpha$
emission-line ratio. Same legend as in Fig.~\ref{diagn}.}

\label{diags}
\end{figure}

The standard diagnostic diagrams are shown in Fig.~\ref{diagn} for
the {[}N\noun{ii}{]} diagnostic and in Fig.~\ref{diags} for the
{[}S\noun{ii}{]} diagnostic. The limit between star-forming galaxies
and AGNs are given by \citet{Kewley:2001ApJS..132...37K} (+ 0.15dex
for the {[}S\noun{ii}{]} diagnostic in order to take into account
the model uncertainties). The star-forming galaxies are well separated
from AGNs and follow a clear sequence covering a large range of ionization
levels and collisional excitation degrees. The classification is obvious,
i.e. the {[}N\noun{ii}{]} and {[}S\noun{ii}{]} diagnostics are
in agreement, for 37 galaxies: 34 are star-forming galaxies and 3
are Seyfert 2.

\subsubsection{{}``Blue'' diagnostic diagrams}

\begin{figure}
\begin{center}\includegraphics[%
  width=1.0\columnwidth,
  keepaspectratio]{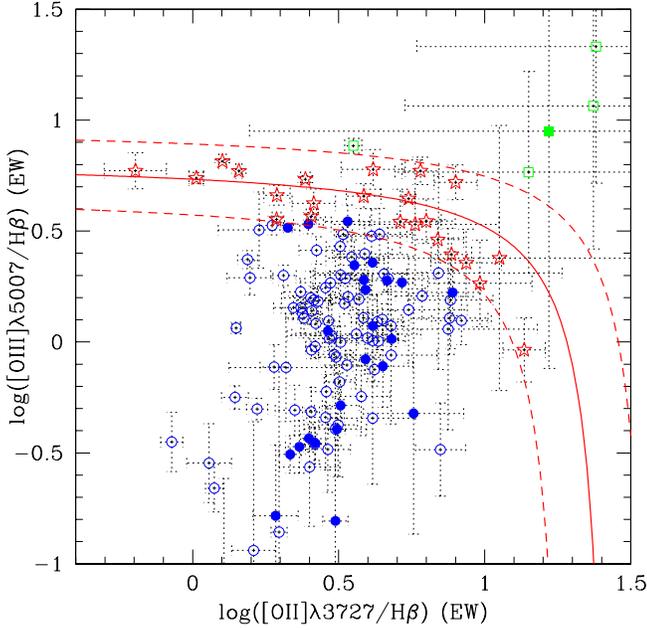}\end{center}

\caption{{}``Blue'' diagnostic diagram. The symbols show the results of
the {}``blue'' diagnostic: blue circles for star-forming galaxies,
green squares for AGNs (Seyfert 2), magenta triangles for contradictory
cases and red stars for unclassified objects. Filled symbols are for
objects already classified with standard diagnostic diagrams. The
solid line shows the empirical calibration from \citet{Lamareille:2004MNRAS.350..396L}
and the dashed lines the associated error domain.}

\label{diagv32}
\end{figure}

\begin{figure}
\begin{center}\includegraphics[%
  width=1.0\columnwidth,
  keepaspectratio]{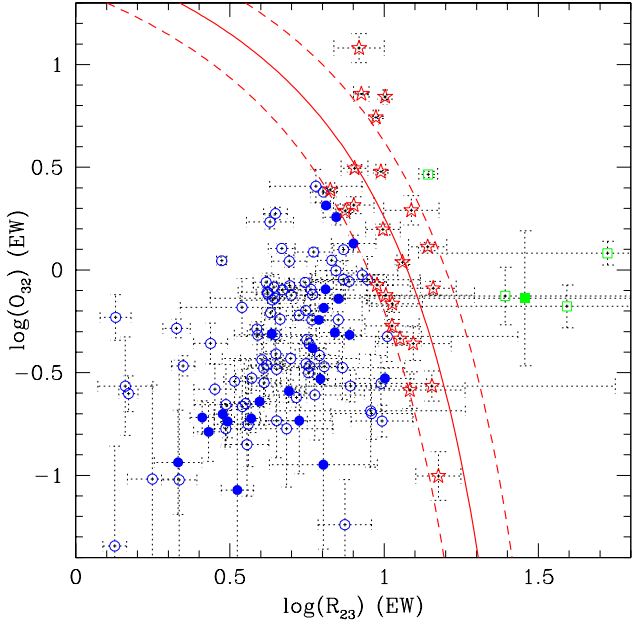}\end{center}

\caption{Another {}``blue'' diagnostic diagram. Same legend as in Fig.~\ref{diagv32}.}

\label{diagr23}
\end{figure}

As recently pointed by \citet{Lamareille:2004MNRAS.350..396L}, we
can also use the {}``blue'' emission lines (i.e. {[}O\noun{ii}{]}$\lambda$3727,
{[}O\noun{iii}{]}$\lambda$5007 and H$\beta$) to perform the spectral
classification for higher redshift galaxies (i.e. with no observable
H$\alpha$ and {[}N\noun{ii}{]}$\lambda$6584 {}``red'' lines)
but with a lower accuracy. We note that 104 galaxies (73.8\% of our
sample) can only be classified with the blue diagnostics. The blue
diagnostic diagrams are shown in Fig.~\ref{diagv32} and in Fig.~\ref{diagr23}.
Please note that we use theses diagrams without any correction for
dust extinction (as calibrated on 2dFGRS data), by the use of equivalent
widths instead of fluxes (see Sect.~\ref{sub:Discussion-on-dust}
below).

We found four objects (\object{LCL05\,045}, \object{LCL05\,065},
\object{LCL05\,109}, and \object{LCL05\,142}), very close in
the $R_{\mathrm{23}}$ vs. $O_{\mathrm{32}}$ classification (four
points on top of Fig.~\ref{diagr23}), which are classified as Seyfert
2 according to this diagram. However this classification is not in
agreement with i) the {}``red'' classification as star-forming galaxy
that we derive for one of them (\object{LCL05\,065}), ii) with
the overall aspect of their spectra (very faint continuum and high
ionization state typical of HII galaxies), or with the low {[}N\noun{ii}{]}/H$\alpha$
ratio estimated recently from NIR spectroscopy for \object{LCL05\,045}
\citep{Maier:2005astro.ph..8239M}. We conclude that the $R_{\mathrm{23}}$
vs. $O_{\mathrm{32}}$ {}``blue'' classification, calibrated on
the 2dFGRS data, may not be valid on its upper part. For these four
objects, we keep only the results from the {[}O\noun{iii}{]}$\lambda$5007/H$\beta$
vs. {[}O\noun{ii}{]}$\lambda$3727/H$\beta$ classification, i.e.
\emph{candidate} star-forming galaxies (see below).

We found four objects (\object{LCL05\,017}, \object{LCL05\,097},
\object{LCL05\,115}, and \object{LCL05\,130}) which are classified
as Seyfert 2 galaxies but with very high error bars on the diagnostic
diagrams. These objects show noisy spectra and/or undetected H$\beta$
emission line (while oxygen lines are detected). Therefore their classification
as Seyfert 2 is not fully secure.

We have a number of objects which fall into the error domain on the
two {}``blue'' diagrams and are thus unclassified. After checking
their spectra, we decided to keep them in our sample as \emph{candidate}
star-forming galaxies, keeping in mind in the subsequent analysis
that their emission-line spectrum could be contaminated by a low-luminosity
AGN.

We finally find 115 (81.6\%) {}``secure'' star-forming galaxies,
7 (5.0\%) Seyfert 2, 16 (11.3\%) {}``candidate'' star-forming galaxies,
and 3 (2.1\%) objects which are still unclassified (i.e. they have
one or more missing lines). Results are shown in Table~\ref{classifdata}.

\subsubsection{Discussion on dust extinction\label{sub:Discussion-on-dust}}

The {}``red'' diagnostic diagram makes use of various line ratios
which are all insensitive to the dust extinction because they involve
emission lines with similar wavelengths ({[}N\noun{ii}{]}$\lambda$6548,
{[}S\noun{ii}{]}$\lambda\lambda$6717,6731 and H$\alpha$ in one
case, {[}O\noun{iii}{]}$\lambda\lambda$4959,5007 and H$\beta$
in the other case). This is not the case for the {}``blue'' diagnostic
diagrams which make use of the {[}O\noun{ii}{]}$\lambda$3727 and
H$\beta$ emission lines in the same ratio. These diagrams can then
be strongly affected by the dust extinction.

The effect of dust is minimized by the use of equivalent width measurements
instead of fluxes. Indeed no correction for reddening is needed on
equivalent width ratios, if we assume that the attenuation in the
continuum and emission lines is the same. To check this assumption,
we have derived dust extinction values from the observed H$\alpha$/H$\beta$
Balmer decrement on the $24$ galaxies where it is possible (we use
the extinction law of \citealp{Seaton:1979MNRAS.187P..73S}, and a
theoretical Balmer decrement of $2.87$ from \citealp{Osterbrock:1989agna.book.....O}).
The E($B-V$) coefficients that we found are given in Table~\ref{classifdata}.
We then used these results to correct the {[}O\noun{ii}{]}$\lambda$3727/H$\beta$
flux ratio for reddening and we compared it to the same equivalent
width ratio.

\begin{figure}
\begin{center}\includegraphics[%
  width=1.0\columnwidth,
  keepaspectratio]{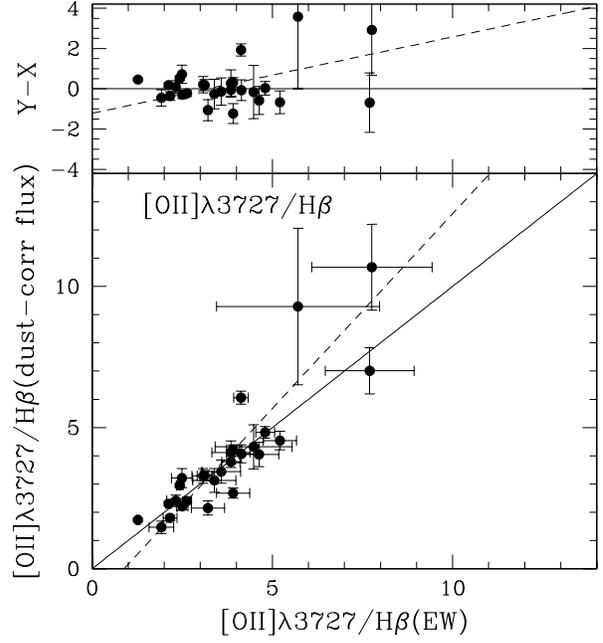}\end{center}

\caption{Comparison between the equivalent-width ($x$-axis) and dust-corrected
flux ($y$-axis) ratio of {[}O\noun{ii}{]}$\lambda$3727 and H$\beta$
emission lines. The solid line shows the $y=x$ curve and the dashed
line is the fit to the data.}

\label{fig:comp_ext}
\end{figure}

Fig.~\ref{fig:comp_ext} shows the result of this comparison. We
see that the equivalent width ratio is consistent with the dust-corrected
flux ratio, with the exception of two very high ratios which are underestimated
with equivalent widths. The rms of the residuals around the $y=x$
line is $0.10$ dex. We conclude that the {}``blue'' diagnostic
diagrams are not significantly affected by the differential attenuation
between {[}O\noun{ii}{]}$\lambda$3727 and H$\beta$ emission lines.
The low value of the rms of the residuals tells us that any possible
effect is already included in the error domain of the {}``blue''
calibration.

\section{Conclusions}

We have defined a sample of 141 emission-line galaxies at intermediate
redshifts ranging from $z=0.2$ to $z=1.0$. We obtained medium-resolution
spectroscopic observations of these galaxies in the optical range,
and associated $R$-band photometry. The following conclusions can
be drawn from this paper:

\begin{itemize}
\item Our sample has been used to test the \texttt{platefit} software originally
developed by C.\ Tremonti, and which is designed to automatically
measure spectral features (e.g. emission lines). We managed to adapt
it to our lower resolution and SNR spectra. The comparison with manual
measurements shows that we get better measurements for those emission
lines where Balmer absorption features are important (e.g. H$\alpha$,
H$\beta$ and {[}O\noun{ii}{]}$\lambda$3727 emission lines), and
that we get correct measurements of flux and equivalent widths for
blended lines (e.g. {[}N\noun{ii}{]}$\lambda$6584 and H$\alpha$
emission lines). 
\item We have done as careful a job as possible and are reasonably sure
that the \texttt{platefit} software can also be used for future and
ongoing large surveys (VVDS, zCOSMOS, etc) which are based on low
resolution spectroscopy. We verify, by downgrading the resolution
of our spectra, that the flux and equivalent-width measurements at
low resolution are not altered more than the measurement error. In
particular, the {[}N\noun{ii}{]}$\lambda$6584/H$\alpha$ line ratio
is robust to resolution changes.
\item The \texttt{platefit} software has been used to measure $k$-corrected
spectroscopic colors. Our sample of galaxies covers all the late-type
range in a color-color diagram, with a maximum for the Irregular and
Sc spectral types. 
\item Standard and {}``blue'' diagnostic diagrams show a majority of star-forming
galaxies, and some narrow-lines AGNs (i.e. Seyfert 2 galaxies), covering
the whole range of ionization level and collisional excitation degree.
Because the H$\alpha$ line gets redshifted out of the optical range
at high redshifts, $\sim$ 70\% of our sample must be classified using
the {}``blue'' diagnostic diagrams. About 10\% of our galaxies still
remain unclassified because they fall in the uncertainty region of
these diagrams, we classify them as {}``candidate'' star-forming
galaxies.
\end{itemize}
More analysis in terms of chemical abundances and stellar populations
will be described in subsequent papers.

\begin{acknowledgements}
We thank C. Tremonti for giving us the right to use the \texttt{platefit}
software. F.L. would like to thank warmly R. Pell\'{o} for decisive
help on photometric reduction and AB correction calculations. J.B.
acknowledges the receipt of an ESA external post-doctoral fellowship.
J.B. acknowledges the receipt of FCT fellowship BPD/14398/2003. We
thank N. Courtney for the photometric calibration of the J1206 field
and R. Ellis for providing us Keck spectroscopy of some magnified
objects. We thank the anonymous referee for useful comments and suggestions.
\end{acknowledgements}
\bibliographystyle{aa}
\bibliography{flamare}

\Online

{\scriptsize
\begin{longtable}{lrrrrrrrrrrrrrr}
\caption{\normalsize
Photometric data. \emph{LCL05}: identification number, see Tables
\ref{cfsample}, \ref{clsample} and \ref{gdsample}.$R_{\mathrm{AB}}$:
photometric magnitude in the $R$ band (or $I$ band for filter 5).
\emph{filter}: the filter used (1: FORS $R$ Gunn, 2: FORS $R$ Bessel,
3: TEK2048 $R$, 4: BTC $R$, 5: CFH12k $I$, 6: WFPC $F606W$). \emph{ape}:
aperture differences (the difference between observed and spectroscopic
magnitude in the given filter). $I_{\mathrm{AB}}^{0}$: $k$-corrected
magnitude in the $I$ band (the sum of the spectroscopic magnitude
in rest-frame and the aperture differences). $A_{I}$: foreground
dust extinction in the $I$ band (rest-frame). $(U-B)_{\mathrm{AB}}^{0}$,$(B-V)_{\mathrm{AB}}^{0}$,
$(V-R)_{\mathrm{AB}}^{0}$, $(V-I)_{\mathrm{AB}}^{0}$: foreground-dust-
and $k$-corrected spectroscopic colors. $(u-g)_{\mathrm{AB}}^{0}$,$(g-r)_{\mathrm{AB}}^{0}$,
$(g-i)_{\mathrm{AB}}^{0}$: foreground-dust- and $k$-corrected spectroscopic
colors with the SDSS filters. $M_{\mathrm{AB}}(I)$: foreground-dust-
and $k$-corrected absolute magnitude in the $I$ band. \emph{lens}:
lensing correction if necessary (already applied to $M_{\mathrm{AB}}(I)$).}
\label{photodata}\\
\hline\hline
LCL05	&$R_{\mathrm{AB}}$	&filter	&	ape&	$I_{\mathrm{AB}}^0$	&$A_{I}$	&$(U-B)_{\mathrm{AB}}^0$	&$(B-V)_{\mathrm{AB}}^0$	&$(V-R)_{\mathrm{AB}}^0$	&$(V-I)_{\mathrm{AB}}^0$	&$(u-g)_{\mathrm{AB}}^0$	&$(g-r)_{\mathrm{AB}}^0$	&$(g-i)_{\mathrm{AB}}^0$	&$M_{\mathrm{AB}}(I)$	&lens	\\
\hline
\endfirsthead
\caption{\normalsize
continued.}\\
\hline\hline
LCL05	&$R_{\mathrm{AB}}$	&filter	&	ape&	$I_{\mathrm{AB}}^0$	&$A_{I}$	&$(U-B)_{\mathrm{AB}}^0$	&$(B-V)_{\mathrm{AB}}^0$	&$(V-R)_{\mathrm{AB}}^0$	&$(V-I)_{\mathrm{AB}}^0$	&$(u-g)_{\mathrm{AB}}^0$	&$(g-r)_{\mathrm{AB}}^0$	&$(g-i)_{\mathrm{AB}}^0$	&$M_{\mathrm{AB}}(I)$	&lens	\\
\hline
\endhead
\hline
\endfoot
001	&$21.22$	&1	&$-0.56$	&$21.03$	&$0.04$	&...	&$0.49$	&$0.17$	&$0.31$	&$0.88$	&$0.31$	&$0.46$	&$-20.26\pm0.04$	&$0.00$	\\
002	&$21.11$	&1	&$-0.54$	&$21.26$	&$0.02$	&$1.01$	&$0.21$	&$0.00$	&$-0.01$	&$0.89$	&$-0.05$	&$-0.07$	&$-21.56\pm0.04$	&$0.00$	\\
003	&$20.47$	&1	&$-0.61$	&$20.37$	&$0.04$	&$1.20$	&$0.31$	&$0.16$	&$0.32$	&$0.68$	&$0.22$	&$0.38$	&$-20.93\pm0.03$	&$0.00$	\\
004	&$19.92$	&1	&$-1.41$	&$19.43$	&$0.04$	&$0.83$	&$0.36$	&$0.28$	&$0.62$	&...	&$0.41$	&$0.69$	&$-21.27\pm0.02$	&$0.00$	\\
005	&$19.19$	&1	&$-0.63$	&$18.68$	&$0.04$	&...	&$0.89$	&$0.30$	&$0.56$	&...	&$0.63$	&$0.92$	&$-22.02\pm0.01$	&$0.00$	\\
006	&$21.14$	&1	&$-0.69$	&$20.96$	&$-0.05$	&$1.14$	&$0.34$	&$0.13$	&$0.26$	&$1.05$	&$0.26$	&$0.37$	&$-19.45\pm0.04$	&$0.00$	\\
007	&$20.15$	&1	&$-0.57$	&$19.85$	&$0.04$	&$0.94$	&$0.58$	&$0.22$	&$0.40$	&...	&$0.47$	&$0.65$	&$-20.86\pm0.02$	&$0.00$	\\
008	&$20.87$	&1	&$-0.61$	&$20.71$	&$0.04$	&...	&$0.30$	&$0.15$	&$0.32$	&...	&$0.30$	&$0.41$	&$-19.76\pm0.03$	&$0.00$	\\
\end{longtable}}

{\scriptsize\begin{longtable}{llllr}
\caption{\normalsize
Spectral classification.\emph{LCL05}: identification number, see Tables~\ref{cfsample},\ref{clsample}~and~\ref{gdsample}.
\emph{Standard}: standard diagnostic. \emph{Blue}: {}``blue'' diagnostic (see text for details). \emph{Type}:
final spectral type. \emph{E($B-V$)}: extinction coefficient computed from H$\alpha$/H$\beta$ ratio.}
\label{classifdata}\\
\hline\hline
LCL05	&Standard	&Blue	&Type	&E($B-V$)	\\
\hline
\endfirsthead
\caption{\normalsize
continued}\\
\hline\hline
LCL05	&Standard	&Blue	&Type	&E($B-V$)	\\
\hline
\endhead
\hline
\endfoot
001	&...	&SF	&Star-forming	&...	\\
002	&...	&SF	&Star-forming	&...	\\
003	&...	&SF	&Star-forming	&...	\\
004	&SF	&SF	&Star-forming	&$0.32\pm0.02$	\\
005	&Sey2	&\textit{undef}	&Seyfert 2	&...	\\
006	&SF	&SF	&Star-forming	&$0.06\pm0.04$	\\
007	&SF	&SF	&Star-forming	&$1.06\pm0.09$	\\
008	&SF	&SF	&Star-forming	&$0.05\pm0.02$	\\
\end{longtable}}

{\scriptsize
\begin{longtable}{lrrrrrrrrrr}
\caption{\normalsize
Measurements of forbidden emission lines. Fluxes are given in $f(H\beta)$
units, associated errors in $\sigma[f(H\beta)]$ units. Equivalent
widths are given in $\AA$.}
\label{datalines1}\\
\hline\hline
	&\multicolumn{2}{r}{[O\sc{ii}]$\lambda$3727}	&\multicolumn{2}{r}{[O\sc{iii}]$\lambda$4959}	&\multicolumn{2}{r}{[O\sc{iii}]$\lambda$5007}	&\multicolumn{2}{r}{[N\sc{ii}]$\lambda$6584}	&\multicolumn{2}{r}{[S\sc{ii}]$\lambda\lambda$6717+6731}	\\
LCL05	&flux&EW	&flux&EW	&flux&EW	&flux&EW	&flux&EW	\\
\hline
\endfirsthead
\caption{\normalsize
continued.}\\
\hline\hline
	&\multicolumn{2}{r}{[O\sc{ii}]$\lambda$3727}	&\multicolumn{2}{r}{[O\sc{iii}]$\lambda$4959}	&\multicolumn{2}{r}{[O\sc{iii}]$\lambda$5007}	&\multicolumn{2}{r}{[N\sc{ii}]$\lambda$6584}	&\multicolumn{2}{r}{[S\sc{ii}]$\lambda\lambda$6717+6731}	\\
LCL05	&flux&EW	&flux&EW	&flux&EW	&flux&EW	&flux&EW	\\
\hline
\endhead
\hline
\endfoot
001	&$2.03\pm1.0$	&$-23.6\pm1.7$	&$0.32\pm0.7$	&$-2.6\pm1.1$	&$0.31\pm0.7$	&$-2.7\pm1.1$	&...	&...	&...	&...	\\
002	&$0.79\pm0.4$	&$-12.4\pm1.1$	&...	&$-0.3\pm1.4$	&$0.25\pm0.5$	&$-3.1\pm1.2$	&...	&...	&...	&...	\\
003	&$2.87\pm1.1$	&$-50.0\pm0.8$	&$0.59\pm0.8$	&$-8.9\pm0.6$	&$1.86\pm0.9$	&$-29.0\pm0.7$	&...	&...	&...	&...	\\
004	&$2.49\pm1.6$	&$-20.5\pm1.0$	&...	&$-0.7\pm1.3$	&$0.16\pm1.2$	&$-1.0\pm0.9$	&$0.91\pm2.9$	&$-6.9\pm1.6$	&$1.50\pm2.8$	&$-13.3\pm2.0$	\\
005	&$1.70\pm1.4$	&$-3.8\pm0.6$	&$0.25\pm1.4$	&$-0.2\pm0.4$	&$5.86\pm1.6$	&$-5.5\pm0.3$	&$2.86\pm3.4$	&$-3.0\pm0.5$	&$3.30\pm3.0$	&$-3.9\pm0.7$	\\
006	&$3.28\pm1.6$	&$-28.5\pm1.7$	&$0.66\pm1.0$	&$-5.3\pm1.2$	&$2.20\pm1.3$	&$-17.6\pm1.3$	&...	&$-1.3\pm2.7$	&$1.01\pm3.7$	&$-14.3\pm6.8$	\\
007	&$3.73\pm1.5$	&$-7.2\pm0.9$	&...	&...	&$0.48\pm1.3$	&$-0.6\pm0.7$	&$2.80\pm2.7$	&$-4.4\pm1.4$	&$1.63\pm2.8$	&$-2.9\pm1.6$	\\
008	&$3.91\pm1.8$	&$-36.3\pm1.3$	&$0.55\pm1.1$	&$-4.7\pm0.9$	&$1.19\pm1.2$	&$-10.4\pm1.0$	&$0.56\pm1.6$	&$-6.7\pm1.7$	&$0.96\pm3.7$	&$-11.3\pm4.2$	\\
\end{longtable}}

{\scriptsize
\begin{longtable}{lrrrrrrrrrrrr}
\caption{\normalsize
Measurements of Balmer emission lines. Fluxes are given in $f(H\beta)$
units, associated errors in $\sigma[f(H\beta)]$ units, $H\beta$
flux and associated error are given in $10^{-17}$ erg s$^{-1}$ cm$^{-2}$
unit. Equivalent widths are given in $\AA$, the \emph{abs} column
gives the equivalent width in absorption.}
\label{datalines2}\\
\hline\hline
	&\multicolumn{3}{r}{H$\alpha$}	&\multicolumn{3}{r}{H$\beta$}	&\multicolumn{3}{r}{H$\gamma$}	&\multicolumn{3}{r}{H$\delta$}	\\
LCL05	&flux&EW&abs	&flux&EW&abs	&flux&EW&abs	&flux&EW&abs	\\
\hline
\endfirsthead
\caption{\normalsize
continued.}\\
\hline\hline
	&\multicolumn{3}{r}{H$\alpha$}	&\multicolumn{3}{r}{H$\beta$}	&\multicolumn{3}{r}{H$\gamma$}	&\multicolumn{3}{r}{H$\delta$}	\\
LCL05	&flux&EW&abs	&flux&EW&abs	&flux&EW&abs	&flux&EW&abs	\\
\hline
\endhead
\hline
\endfoot
001	&...	&...	&$0.0$	&$5.73\pm0.8$	&$-8.1\pm1.3$	&$3.7$	&$0.25\pm0.9$	&$-2.0\pm1.3$	&$2.6$	&$0.51\pm0.9$	&$-3.7\pm1.2$	&$4.3$	\\
002	&...	&...	&$0.0$	&$11.79\pm1.7$	&$-10.9\pm1.7$	&$1.6$	&$0.25\pm0.4$	&$-3.2\pm0.9$	&$2.8$	&$0.12\pm0.3$	&$-1.4\pm0.8$	&$4.4$	\\
003	&...	&...	&$0.0$	&$19.27\pm0.8$	&$-15.0\pm0.7$	&$4.6$	&$0.42\pm0.9$	&$-5.6\pm0.7$	&$3.7$	&$0.32\pm0.8$	&$-3.9\pm0.6$	&$4.4$	\\
004	&$4.08\pm2.2$	&$-29.8\pm1.5$	&$2.3$	&$6.97\pm0.5$	&$-6.7\pm0.6$	&$3.7$	&$0.59\pm0.9$	&$-3.5\pm0.5$	&$2.8$	&$0.52\pm1.0$	&$-2.9\pm0.5$	&$4.8$	\\
005	&$1.18\pm1.6$	&$-1.2\pm0.4$	&$1.0$	&$3.84\pm0.5$	&$-0.9\pm0.2$	&$1.7$	&$0.91\pm0.9$	&$-1.1\pm0.2$	&$0.1$	&$0.40\pm0.9$	&$-0.6\pm0.3$	&$0.6$	\\
006	&$3.05\pm1.8$	&$-39.9\pm2.5$	&$3.1$	&$5.36\pm0.6$	&$-8.0\pm1.1$	&$1.7$	&$0.42\pm0.9$	&$-3.0\pm1.0$	&$3.2$	&$0.44\pm0.9$	&$-3.3\pm1.0$	&$2.5$	\\
007	&$9.11\pm2.2$	&$-14.0\pm1.3$	&$3.3$	&$2.32\pm0.6$	&$-1.3\pm0.5$	&$2.3$	&$0.19\pm0.9$	&$-0.3\pm0.5$	&$1.5$	&$1.02\pm1.0$	&$-1.4\pm0.5$	&$3.4$	\\
008	&$3.02\pm1.8$	&$-36.2\pm1.5$	&$2.8$	&$8.02\pm0.5$	&$-8.8\pm0.8$	&$1.9$	&$0.32\pm0.9$	&$-2.5\pm0.7$	&$2.0$	&$0.24\pm0.8$	&$-1.8\pm0.7$	&$3.5$	\\
\end{longtable}}

\end{document}